\RequirePackage{pdf14}
\documentclass[11pt]{article}
\usepackage{graphicx}
\usepackage{amssymb}
\usepackage{bigints}
\usepackage{tikz}
\usetikzlibrary{fit,positioning,calc}
\usepackage{algorithm}
\usepackage{algorithmic}
\usepackage{float}
\usepackage{subfigure}
\usepackage[default]{jasa_harvard}    
\usepackage{JASA_manu}
\usepackage{cleveref}

\renewcommand{\baselinestretch}{1.35} 

\newtheorem{theorem}{Theorem}[section]
\newtheorem{lemma}[theorem]{Lemma}
\newtheorem{proposition}[theorem]{Proposition}

\newenvironment{proof}[1][Proof]{\begin{trivlist}
\item[\hskip \labelsep {\bfseries #1}]}{\end{trivlist}}

\newcommand{\qed}{\nobreak \ifvmode \relax \else
      \ifdim\lastskip<1.5em \hskip-\lastskip
      \hskip1.5em plus0em minus0.5em \fi \nobreak
      \vrule height0.75em width0.5em depth0.25em\fi}
\makeatletter
\newenvironment{breakablealgorithm}
  {
   \begin{center}
     \refstepcounter{algorithm}
     \hrule height.8pt depth0pt \kern2pt
     \renewcommand{\caption}[2][\relax]{
       {\raggedright\textbf{\ALG@name~\thealgorithm} ##2\par}%
       \ifx\relax##1\relax 
         \addcontentsline{loa}{algorithm}{\protect\numberline{\thealgorithm}##2}%
       \else 
         \addcontentsline{loa}{algorithm}{\protect\numberline{\thealgorithm}##1}%
       \fi
       \kern2pt\hrule\kern2pt
     }
  }{
     \kern2pt\hrule\relax
   \end{center}
  }
\makeatother

\def\T{{ \mathrm{\scriptscriptstyle T} }}
\begin{document}

\title{\vspace{-50pt}
\date{}
Nonparametric Bayes Modeling of Populations of Networks}
\author{Daniele Durante$^{\ast}$, David B. Dunson$^{\dag}$ and  Joshua T. Vogelstein$^{\star}$\\ \vspace{-5pt}
\\ $^{\ast}$Department of Statistical Sciences, University of Padova, Padova, Italy \\
$^{\dag}$Department of Statistical Science, Duke University, Durham, NC, USA  \\ 
$^{\star}$Department of Biomedical Engineering and  Institute for Computational Medicine,\\ Johns Hopkins University, Baltimore, MD, USA; Child Mind Institute, New York, NY, USA \\ \vspace{-5pt}
\\ e-mail: $^{\ast}$\texttt{durante@stat.unipd.it},  $^{\dag}$\texttt{dunson@stat.duke.edu}, $^{\star}$\texttt{jovo@jhu.edu}}
\maketitle

\vspace*{-.3in}


\begin{center}
\textbf{Abstract}
\end{center}
Replicated network data are increasingly available in many research fields. In connectomic applications, inter-connections among brain regions are collected for each patient under study, motivating statistical models which can flexibly characterize the probabilistic generative mechanism underlying these network-valued data. Available models for a single network are not designed specifically for inference on the entire probability mass function of a network-valued random variable and therefore lack flexibility in characterizing the distribution of relevant topological structures. We propose a flexible Bayesian nonparametric approach for modeling the population distribution of network-valued data.  The joint distribution of the edges is defined via a mixture model which reduces dimensionality and efficiently incorporates network information within each mixture component by leveraging latent space representations. The formulation leads to an efficient Gibbs sampler and provides simple and coherent strategies for inference and goodness-of-fit assessments. We provide theoretical results on the flexibility of our model and illustrate improved performance --- compared to state-of-the-art models --- in simulations and application to human brain networks.
\vspace*{.1in}


\section{Introduction}
There is an increasing availability of object-type data in many fields of application covering biology \cite{ull_2013}, social science \cite{kold_2009} and machine learning \cite**{prats_2011} --- among others. These data are typically generated from random variables defined on non-standard spaces and therefore require adaptation of classical modeling frameworks  in order to ensure meaningful and robust inference.  The importance of this endeavor has motivated a growing  interest towards statistical models, which are sufficiently flexible in characterizing the probabilistic generative mechanism associated with these object-type random variables, and avoid issues arising from model miss-specification. 

Although key improvements have been made for a broad variety of object-type data covering functions \cite{ram_2005}, tensors \cite{dun_2009}, shapes \cite{bhatta_2010} and trees \cite{wang_2007} --- among others --- similarly flexible procedures for network-valued data measuring interconnections among a set of nodes are currently lacking. A main reason for this delay is that the routine collection of such data is a recent development, so that the rich literature on modeling of a single network has not had the opportunity to catch up with the increasing availability of replicated networks. We aim to address this gap by developing a parsimonious and flexible representation of the distribution associated to a network-valued random variable. In accomplishing this goal, we propose a statistical model which reduces dimensionality and characterizes a broad variety of generative mechanisms for network data. 

\begin{figure}[t]
\centering
\includegraphics[height=7cm, width=14cm]{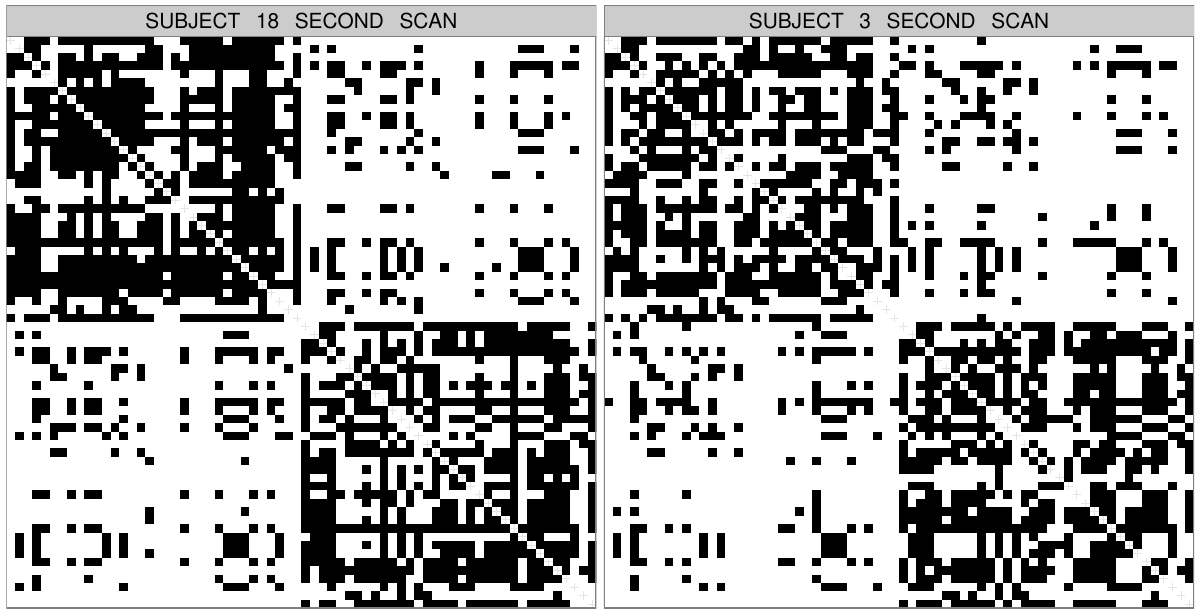}
\caption{\footnotesize{Adjacency matrices representing selected brain networks. Black refers to an edge; white to a non-edge.}}
\label{F1}
\end{figure}

We are motivated by neuroscience studies of brain activity networks --- connectomes --- providing data on the undirected structural interconnections among anatomical brain regions for multiple subjects in a study. Data consist of a collection of $V \times V$  symmetric binary adjacency matrices $\boldsymbol{A}_1, \ldots, \boldsymbol{A}_n$, having elements ${A}_{i[vu]}={A}_{i[uv]}=1$ if brain regions $v=2, \ldots, V$ and $u=1, \ldots, v-1$ are connected by at least one white matter fiber in subject $i=1, \ldots, n$, and ${A}_{i[vu]}={A}_{i[uv]}=0$ if no white matter fibers are detected. In our application, these regions correspond to those defined in the Desikan atlas \cite**{des_2006}, for a total of $V=68$ nodes equally divided in the left and right hemisphere. Data for selected brains are represented in Figure \ref{F1}. 

A primary focus in the literature is on inferring common network patterns and topological properties, such as small-world, scale free and community structures; see e.g. \citeasnoun{bull_2009},  \citeasnoun{bull_2012} and  \citeasnoun{stam_2014}.  Current practice either conducts separate analyses for each $\boldsymbol{A}_i$ to extract network measures \cite**{hag_2008} or applies standard network analyses after averaging $\boldsymbol{A}_1,\ldots,\boldsymbol{A}_n$ \cite{sche_2010}.  We consider instead the replicated networks $\boldsymbol{A}_1, \ldots, \boldsymbol{A}_n$ as realizations from a common network-valued random variable and focus inference on the unknown population distribution associated to this variable.

The literature has essentially ignored the problem of nonparametric inference on the distribution of network-valued data, although there is a vast literature on modeling of a single network $\boldsymbol{A}$; see \citeasnoun{sch_2013} for a review.  Some canonical approaches include exponential random graphs  \cite{fra_1986}, stochastic block models \cite{now_2001},  mixed membership stochastic block models \cite**{air_2008} and latent space models \cite**{hof_2002}.  Our interest is in extending such models to allow replicated networks from a common population distribution, facilitating robust inference on expected network properties and coherently estimating the entire distribution of these topological characteristics across replicates. In effectively accomplishing this goal, we seek a provably flexible representation of the population distribution.

With this motivation, we focus in particular on latent space models, which define the edges of $\boldsymbol{A}$ as conditionally independent given their corresponding edge probabilities, with these probabilities defined as a function of pairwise Euclidean distances between the nodes in a latent space \cite{hof_2002}.  Such models can 
accommodate community behaviors, transitive relations, and $k$-star structures \cite{hof_2002}, with generalizations capturing additional network properties \cite**{kri_2009}, accounting for different types of distance  \cite{hoff_2005,hoff_2007} and improving modeling of the node-specific latent coordinates via hierarchical specifications \cite**{hand_2007}.  There is additionally a literature on multiview networks, focusing on joint modeling of views $\boldsymbol{A}_1,\ldots,\boldsymbol{A}_n$, encoding relationships among a common set of nodes in $n$ different contexts  \cite{gol_2013}.  Multiview network data can be structured as a single three-way tensor, with the first two dimensions indexing the different nodes and the third dimension indexing the context.  Such data differ fundamentally from replicated network data $\boldsymbol{A}_i$ sampled from a common population distribution, for each $i=1, \ldots, n$.  

In this article, we build on latent space models using a Bayesian nonparametric approach to induce a prior on the unknown population distribution.  As each $\boldsymbol{A}_i$ consists of binary elements, we can characterize the population distribution through an unknown probability mass function (pmf) for a network-valued random variable. This probability mass function is characterized using a mixture of low-rank factorizations leveraging latent space representations.  Placing priors on the terms in the factorization, we induce a provably flexible prior for the probability mass function, which leads to a simple approach to posterior computation and facilitates tractable inference procedures.

In order to provide useful insights on network properties and topological structures, we exploit the posterior distribution for the networks' probability mass function to draw meaningful inference on (a) the induced posterior distributions for the expectation of selected network summary measures and (b) the entire distribution of these summary measures characterizing differences in network properties across replicates. We show how a provably flexible model for the entire probability mass function of a network-valued random variable can substantially improve performance in addressing (a) and (b).

We describe our proposed model in Section 2, discussing properties and interpretation.  Section 3 focuses on prior specification, while Section 4 outlines steps for posterior computation and goodness-of-fit assessments via posterior predictive checks.  Section 5 contains a simulation study demonstrating substantial performance gains over competitors, and Section 6 applies the methods to human brain network data.  Section 7 contains concluding remarks.  

\section{Nonparametric Model}
\subsection{Notation and motivation}
Let $\boldsymbol{A}_1,\ldots, \boldsymbol{A}_n$ denote multiple observations of an undirected network with no self-relations and node set $\mathbb{V}$ with cardinality $|\mathbb{V}|=V.$ Each observation $\boldsymbol{A}_{i}$ corresponds to a symmetric $V \times V$ adjacency matrix with binary entries $A_{i[vu]}=A_{i[uv]} \in \{0, 1 \}$ encoding the presence or absence of an edge among nodes $v$ and $u$ for unit $i$.  As self-relationships are not of interest and $\boldsymbol{A}_i$ is symmetric, statistical modeling of $\boldsymbol{A}_1,\ldots, \boldsymbol{A}_n$ coincides with defining a probabilistic generative mechanism underlying data $\mathcal{L}(\boldsymbol{A}_1), \ldots, \mathcal{L}(\boldsymbol{A}_n)$, with  $\mathcal{L}(\boldsymbol{A}_i)=\left(A_{i[21]}, A_{i[31]}, \ldots, A_{i[V1]}, A_{i[32]}, \ldots,  A_{i[V2]}, \ldots, A_{i[V(V-1)]}\right)^\T \in \mathbb{A}_{V}$, the vector encoding the lower triangular elements of $\boldsymbol{A}_i$, for each $i=1, \ldots, n$. 

Data $ \mathcal{L}(\boldsymbol{A}_1), \ldots, \mathcal{L}(\boldsymbol{A}_n)$ are realizations from a multivariate random variable $\mathcal{L}(\boldsymbol{\mathcal{A}})$ with binary entries $\mathcal{L}({\mathcal{A}})_l \in \{0,1\}$, measuring presence or absence of an edge among each pair of nodes $l=1, \ldots, V(V-1)/2$. 
Since there are finitely many network configurations, $\mathcal{L}(\boldsymbol{\mathcal{A}})$ can be seen as a categorical random variable with each category representing one of the possible network configurations $\boldsymbol{a} \in \mathbb{A}_V= \{0,1\}^{V(V-1)/2}$. Considering for example $V=3$, the network-valued random variable  $\mathcal{L}(\boldsymbol{\mathcal{A}})$ has $2^{V(V-1)/2}=8$ possible network configurations $\{(0,0,0); (1,0,0); \ldots; (1,1,1)\}$ and $2^{V(V-1)/2}-1=7$ parameters are required to fully characterize the pmf $p_{ \mathcal{L}(\boldsymbol{\mathcal{A}})}(\boldsymbol{a})=\mbox{pr}\{ \mathcal{L}(\boldsymbol{\mathcal{A}})= \boldsymbol{{a}}\}$, $\boldsymbol{{a}} \in \mathbb{A}_V$ under the restriction $\sum_{\boldsymbol{{a}} \in  \mathbb{A}_V} p_{ \mathcal{L}(\boldsymbol{\mathcal{A}})}(\boldsymbol{a})=1$. 

The number of parameters is intractable and massively larger than the sample size $n$ even in small $V$ settings. In the motivating neuroscience study, brain images have been processed to obtain adjacency matrices considering $V=68$ brain regions.  This implies that, in the absence of constraints, there are  $2^{68\cdot 67/2}-1=2^{2278}-1$ free parameters to estimate characterizing $p_{ \mathcal{L}(\boldsymbol{\mathcal{A}})}$.  Clearly no studies will ever have this many subjects, and hence it is necessary to substantially reduce dimensionality to make the problem tractable.  However, in reducing dimension, it is important to avoid restrictive assumptions that lead to inadequate characterization of the observed network data.

One possibility  is to rely on nonparametric tensor factorization models for multivariate categorical data  \cite**{dun_2009}. However, as the length of the binary vectors of edges is quadratic in the number of nodes, models that do not exploit the network structure for dimensionality reduction are expected to have poor performance when the number of nodes is moderate to large.  In fact, the key difference between a network-valued random variable and an unstructured categorical random vector is that the network configurations share a common underlying structure which informs edge probabilities. Hence, by carefully accommodating network structure  in modeling of $ \mathcal{L}(\boldsymbol{A}_1), \ldots, \mathcal{L}(\boldsymbol{A}_n)$, one might efficiently borrow information across units and within each network, facilitating dimensionality reduction without substantially affecting flexibility.

\subsection{Model formulation and properties}
Motivated by the above discussion, we consider a representation in which $p_{ \mathcal{L}(\boldsymbol{\mathcal{A}})}$  is assigned a mixture model, improving flexibility and facilitating borrowing of information across replicates.  Within each mixture component the edges are defined as conditionally independent Bernoulli random variables given their corresponding component-specific edge probabilities, obtaining
\begin{eqnarray}
p_{\mathcal{L}(\boldsymbol{\mathcal{A}})}(\boldsymbol{a})=\mbox{pr}\{ \mathcal{L}(\boldsymbol{\mathcal{A}})=\boldsymbol{a}\}=  \sum_{h=1}^{H} \nu_h \prod_{l=1}^{V(V-1)/2} (\pi_{l}^{(h)})^{a_l} (1-  \pi^{(h)}_{l})^{1-a_l},
\label{eq1}
\end{eqnarray}
for every network configuration $\boldsymbol{a} \in \mathbb{A}_V$, with $\nu_h \in (0,1)$ the probability assigned to mixture component $h$, for every $h=1, \ldots, H$ and  $ \pi^{(h)}_{l}\in (0,1)$ the probability that an edge is observed for the $l$th pair of nodes in mixture component $h$, for every $h=1, \ldots, H$ and $l=1, \ldots, V(V-1)/2$. In order to incorporate network information and reduce dimensionality within each mixture component, we exploit latent space characterizations of network data and consider a structured representation of each component-specific edge probability vector $\boldsymbol{\pi}^{(h)}=(\pi^{(h)}_{1}, \ldots, \pi^{(h)}_{V(V-1)/2} )^\T \in (0,1)^{V(V-1)/2}$, for $h=1, \ldots, H$. In particular, each $\boldsymbol{\pi}^{(h)}$ is defined as a function of a shared similarity vector $\boldsymbol{Z} \in \Re^{V(V-1)/2}$ and component-specific deviation $\boldsymbol{D}^{(h)} \in \Re^{V(V-1)/2}$ characterized via matrix factorization representations:
\begin{eqnarray}
\boldsymbol{\pi}^{(h)}=\left\{1+\exp({-\boldsymbol{Z}-\boldsymbol{D}^{(h)}})\right\}^{-1},\quad \boldsymbol{D}^{(h)}=\mathcal{L}(\boldsymbol{X}^{(h)}\boldsymbol{\Lambda}^{(h)} \boldsymbol{X}^{(h)\T}), \quad h=1, \ldots, H, 
\label{eq1.pi}
\end{eqnarray}
where $\boldsymbol{X}^{(h)} \in \Re^{V\times R}$ is a $V\times R$ matrix whose rows ${\bf X}_v^{(h)\T}=(X_{v1}^{(h)}, \ldots, X_{vR}^{(h)} ) \in \Re^{R}$, $v=1, \ldots, V$, denote the $R$ latent coordinates of each node $v=1, \ldots, V$ in component $h$, while $\boldsymbol{\Lambda}^{(h)}$ is a $R\times R$ diagonal matrix whose diagonal elements $\lambda^{(h)}_{r} \geq 0$, $r=1, \ldots,  R$ measure the importance of each dimension $r=1, \ldots, R$ in defining   $\boldsymbol{D}^{(h)}$ via \eqref{eq1.pi} in component $h$. In representation \eqref{eq1.pi} the logistic mapping is applied element-wise.

According to representation  \eqref{eq1.pi}, the probability $\pi^{(h)}_l$ of an edge between the $l$th pair of nodes in component $h$ increases with $Z_l$ and $D_l^{(h)}$.  The shared similarities $Z_l$ facilitate centering the different mixture components and improve computational performance. These quantities are modeled as unstructured.  By borrowing information across all networks in all mixture components, we can accurately infer $\boldsymbol{Z}=(Z_1, \ldots, Z_{V(V-1)/2})^{\T}$ without additional structural constraints in our experience.  There is much less information in the data about the component-specific deviations $D_l^{(h)}$, and we rely on a low-rank matrix factorization as in equation \eqref{eq1.pi}. In particular --- letting $l$ denote the pair of nodes $v$ and $u$, with $v>u$ --- we define $D_l^{(h)}=\mathcal{L}({X}^{(h)}{\Lambda}^{(h)} {X}^{(h)\T})_l={\bf X}_v^{(h)\T}{\Lambda}^{(h)}{\bf X}_u^{(h)}=\sum_{r=1}^R \lambda^{(h)}_r{X}_{vr}^{(h)}{X}_{ur}^{(h)}$. 

Beside reducing dimensionality, the above weighted dot product representation has an appealing interpretation. Recalling our motivating neuroscience application and focusing on mixture component $h$, the coordinate $X^{(h)}_{vr} \in \Re$ may measure the activity of brain region $v$ within pathway $r$, for each $v=1, \ldots, V$ and $r=1, \ldots, R$. According to the dot product construction, regions with activities in the same direction --- both positive or negative --- will be more similar. The similarity --- or dissimilarity --- will be higher the stronger the activity is in the same --- or opposite --- direction, with  $\lambda^{(h)}_{r} \geq 0$ measuring the importance of each pathway $r=1, \ldots, R$. The closer is $\lambda^{(h)}_{r}$ to zero, the lower the contribution of pathway $r$ in defining $\boldsymbol{D}^{(h)}$ via \eqref{eq1.pi} in component $h$.

Factorization \eqref{eq1.pi} for $\boldsymbol{\pi}^{(h)}$, $h=1, \ldots, H$, adapts dot product characterizations of edge probabilities for a single network \cite{hoff_2005,hoff_2007} to a mixture representation. In latent space modeling of a single network, this representation has been shown to provide a more general characterization of interconnection structures and network properties than  stochastic block models \cite{now_2001},  mixed membership stochastic block models \cite**{air_2008} and latent distance models \cite**{hof_2002}. However, these procedures have the disadvantage of  characterizing the observed data through a single  edge probability vector, which forces $p_{\mathcal{L}(\boldsymbol{\mathcal{A}})}$ to concentrate its mass on a subset of configurations characterized by a specific network property, while ruling out others. 

Network properties and topological structures can vary substantially, with some subsets of the data having small-world behaviors, while others indicate strong community patterns --- for example.  Model \eqref{eq1}--\eqref{eq1.pi} adaptively assigns probability to different subsets of configurations, each one potentially characterized by a different network property, substantially improving flexibility, while considerably reducing the dimensionality from $2^{V(V-1)/2}-1$ to $H\{1+R(V+1)\}+V(V-1)/2-1$ parameters. As formalized in Lemma \ref{theorem1}, our mixture of low-rank factorizations can represent any possible pmf $p_{\mathcal{L}(\boldsymbol{\mathcal{A}})} \in  \mathcal{P}_{2^{V(V-1)/2}}$ defined on a network-valued sample space.  
\begin{lemma}
Any $p_{\mathcal{L}(\boldsymbol{\mathcal{A}})} \in  \mathcal{P}_{2^{V(V-1)/2}}$ admits representation (\ref{eq1}) for some $H$ with  $\nu_h \in (0,1)$ mixing weights such that  $\sum_{h=1}^{H} \nu_h=1$ and each $\boldsymbol{\pi}^{(h)} \in (0,1)^{V(V-1)/2}$ factorized as in \eqref{eq1.pi} for some $R$.
\label{theorem1}
\end{lemma}

This confirms the full flexibility of our construction, which can be viewed as nonparametric given appropriately chosen priors for the components.
All proofs can be found in the Appendix.

\subsection{Identifiability and inference}
Factorization \eqref{eq1.pi} is not unique.  For example, letting $\boldsymbol{\tilde{Z}}=\boldsymbol{Z}+\boldsymbol{U}$ and $\boldsymbol{\tilde{D}}^{(h)}=\boldsymbol{{D}}^{(h)}-\boldsymbol{U}$, for each $h=1, \ldots, H$ then $\boldsymbol{\tilde{Z}}+\boldsymbol{\tilde{D}}^{(h)}=\boldsymbol{Z}+\boldsymbol{U}+\boldsymbol{{D}}^{(h)}-\boldsymbol{U}=\boldsymbol{Z}+\boldsymbol{{D}}^{(h)}$. This further affects the uniqueness of the factorization $\boldsymbol{D}^{(h)}=\mathcal{L}(\boldsymbol{X}^{(h)}\boldsymbol{\Lambda}^{(h)} \boldsymbol{X}^{(h)\T})$. Moreover, there exist infinitely many diagonalizable positive semidefinite matrices having $\boldsymbol{D}^{(h)}$ as lower triangular elements.  Similar issues arise routinely in Bayesian factorizations and nonparametric models, which tend to be purposely over-complete.  Such over-completeness often has a beneficial effect on computational efficiency and does not lead to problems  
when inference focuses on identifiable functionals of the parameters.  Refer for example to \citeasnoun{ghosh_2009} and \citeasnoun{bha_2011}. 

A main focus of inference in network analysis is on studying selected network summary measures of interest, covering network density, transitivity and average path length  --- among others --- in order to understand whether the generative mechanism underlying the observed data is characterized by particular topological structures and how these properties are distributed across replicates. Letting $\theta_k=g_k\{ \mathcal{L}(\boldsymbol{\mathcal{A}})\}$ the random variable associated with the $k$th network summary measure, for each $k=1, \ldots, K$, we allow inference on (a) expected network measures $\mbox{E}(\theta_k)=\bar{\theta}_k$, $k=1, \ldots, K$ and (b) the entire distribution $p_{\theta_k}(\theta)$, $k=1, \ldots, K$ of these measures across replicates. Both quantities are explicitly available as functionals of  $p_{\mathcal{L}(\boldsymbol{\mathcal{A}})}$ via $\mbox{E}(\theta_k)=\bar{\theta}_k=\sum_{\boldsymbol{{a}} \in \mathbb{A}_V}g_k(\boldsymbol{{a}})p_{\mathcal{L}(\boldsymbol{\mathcal{A}})}(\boldsymbol{{a}})$ and $p_{\theta_k}(\theta)=\sum_{\boldsymbol{{a}} \in \mathbb{A}_V: g_k(\boldsymbol{{a}})= \theta}p_{\mathcal{L}(\boldsymbol{\mathcal{A}})}(\boldsymbol{{a}})$, $k=1, \ldots, K$, respectively. Therefore, we follow \citeasnoun{ghosh_2009} in avoiding identifiability constraints in factorization  \eqref{eq1.pi} for the edge probability vectors  $\boldsymbol{\pi}^{(h)}$, $h=1, \ldots, H$, as they are not required to ensure identifiability of $p_{\mathcal{L}(\boldsymbol{\mathcal{A}})}$.

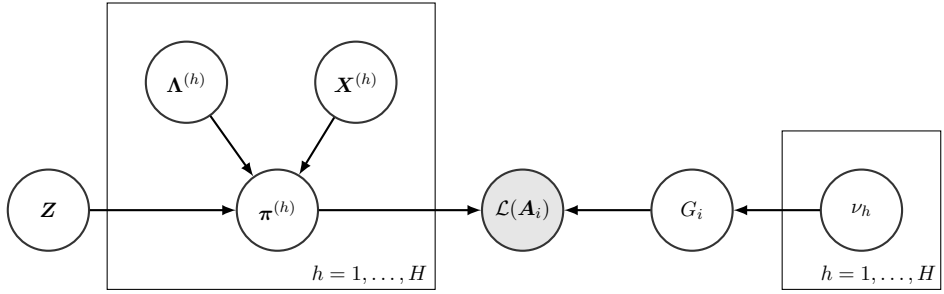
\begin{figure}
\centering
\begin{tikzpicture}[scale=0.72, transform shape]
\tikzstyle{main}=[circle, minimum size = 15mm, thick, draw =black!80, node distance = 16mm]
\tikzstyle{connect}=[-latex, thick]
\tikzstyle{box}=[rectangle, draw=black!100]
  \node[main, fill = white!100] (pi) {$\boldsymbol{\pi}^{(h)}$ };
    \node[main, fill = black!10] (A) [right= 3.0cm of pi] {$\mathcal{L}(\boldsymbol{A}_i)$ };
  \node[main] (G) [right= of A] {$G_i$};
    \node[main] (nu) [right=of G] {$\nu_h$};
  \node[main] (x) [above= of {$(pi)!0.32!(A)$}]  {$\boldsymbol{X}^{(h)}$};
            \node[main] (lam) [left= of x]  {$\boldsymbol{\Lambda}^{(h)}$};
             \node[main] (Z) [left= 2.7cm of pi]  {$\boldsymbol{Z}$};
               \path        (G) edge [connect] (A);
\path        (nu) edge [connect] (G);
\path        (pi) edge [connect] (A);
\path        (x) edge [connect] (pi);
\path        (lam) edge [connect] (pi);
\path        (Z) edge [connect] (pi);
    \node[rectangle, inner sep=7mm, draw=black!100, fit = (pi) (lam) (x) (lam)] {};
      \node[rectangle, inner sep=7mm, draw=black!100, fit = (nu)] {};
       \node[] at (11.1,-1.2) {$h=1, \ldots, H$};
                \node[] at (1.7,-1.2) {$h=1, \ldots, H$};
 \end{tikzpicture}
\caption{{\footnotesize{Graphical representation of the probabilistic mechanism generating  data $\mathcal{L}(\boldsymbol{A}_i)$ from  \eqref{eq1}--\eqref{eq1.pi}.}}}
\label{F_mix}
\end{figure}

Our formulation  \eqref{eq1}--\eqref{eq1.pi} additionally facilitates Monte Carlo integration methods for inference on quantities (a) and (b). According to Figure \ref{F_mix}, generating a  network $\mathcal{L}(\boldsymbol{A}_i)$  from our statistical model  \eqref{eq1}--\eqref{eq1.pi} first relies on sampling a component indicator variable $G_i \in \{1, \ldots, H \}$ with pmf defined by the mixing probabilities $\nu_1, \ldots, \nu_H$, so that $\mbox{pr}(G_i=h)=\nu_h$, for each $h=1, \ldots, H$. Conditionally on $G_i=h$ and given the edge probability vector $\boldsymbol{\pi}^{(h)}$ in component $h$ --- factorized according to \eqref{eq1.pi} --- the network $\mathcal{L}(\boldsymbol{A}_i)$  is generated by sampling its edges $\mathcal{L}({A}_i)_l$ from conditionally independent Bernoulli random variables with probabilities ${\pi}_l^{(h)}$, for each $l=1, \ldots, V(V-1)/2$.

Beside studying selected network summary measures, it is additionally interesting to consider the expectation of the network-valued random variable, which summarizes an averaged network architecture. This quantity is easily available as $\mbox{E}\{\mathcal{L}(\boldsymbol{\mathcal{A}}) \}=\bar{\boldsymbol{\pi}}=\sum_{\boldsymbol{{a}} \in  \mathbb{A}_V} \boldsymbol{a} p_{ \mathcal{L}(\boldsymbol{\mathcal{A}})}(\boldsymbol{a})$,
and --- according to Proposition \ref{theorem2} ---  can be analytically computed as the weighted sum of the edge probability vectors $\boldsymbol{\pi}^{(1)}, \ldots, \boldsymbol{\pi}^{(H)}$, with weights given by $\nu_1,\ldots, \nu_H$.
\begin{proposition}
Under representation \eqref{eq1} for $p_{ \mathcal{L}(\boldsymbol{\mathcal{A}})}$, the expected value for the network-valued random variable $\mathcal{L}(\boldsymbol{\mathcal{A}})$ is given by $\bar{\boldsymbol{\pi}}=\mbox{E}\{\mathcal{L}(\boldsymbol{\mathcal{A}}) \}=\sum_{\boldsymbol{{a}} \in  \mathbb{A}_V} \boldsymbol{a} p_{ \mathcal{L}(\boldsymbol{\mathcal{A}})}(\boldsymbol{a})=\sum_{h=1}^H \nu_h \boldsymbol{\pi}^{(h)}$.
\label{theorem2}
\end{proposition}

\section{Prior Specification and Properties}
Results in Section 2 ensure that any true probability mass function for a population of networks $p^0_{ \mathcal{L}(\boldsymbol{\mathcal{A}})} \in  \mathcal{P}_{2^{V(V-1)/2}}$ admits representation (\ref{eq1}), with component-specific edge probability vectors $\boldsymbol{\pi}^{(h)}$ factorized as in (\ref{eq1.pi}). Although this is a key result, it is not guaranteed that the same flexibility is maintained after choosing independent priors $\boldsymbol{Z} \sim \Pi_{Z}$, $\boldsymbol{\nu}=(\nu_1, \ldots, \nu_H) \sim \Pi_{\nu}$, $\boldsymbol{X}^{(h)}\sim \Pi_{X}$ and $\boldsymbol{\lambda}^{(h)}=(\lambda^{(h)}_1, \ldots, \lambda^{(h)}_R)^\T\sim\Pi_{\lambda}$, for  $h=1, \ldots, H$. 

Letting $\mathbb{B}_{\epsilon}(p^0_{ \mathcal{L}(\boldsymbol{\mathcal{A}})})=\{p_{ \mathcal{L}(\boldsymbol{\mathcal{A}})}: \sum_{\boldsymbol{a} \in \mathbb{A}_V} |p_{ \mathcal{L}(\boldsymbol{\mathcal{A}})}(\boldsymbol{a})- p^0_{ \mathcal{L}(\boldsymbol{\mathcal{A}})}(\boldsymbol{a})|<\epsilon \}$ denote an $L_1$ neighborhood around any  $p^0_{ \mathcal{L}(\boldsymbol{\mathcal{A}})} \in  \mathcal{P}_{2^{V(V-1)/2}}$, we place simple conditions on $\Pi_{Z}$, $\Pi_{\nu}$, $\Pi_{X}$ and $\Pi_{\lambda}$, so that the prior $\Pi$ on $p_{ \mathcal{L}(\boldsymbol{\mathcal{A}})}$ induced through (\ref{eq1})--(\ref{eq1.pi}) has full support on $\mathcal{P}_{2^{V(V-1)/2}}$, meaning that  $\Pi \{ \mathbb{B}_{\epsilon}(p^0_{ \mathcal{L}(\boldsymbol{\mathcal{A}})})\}>0$ for any  $p^0_{ \mathcal{L}(\boldsymbol{\mathcal{A}})} \in  \mathcal{P}_{2^{V(V-1)/2}}$ and $\epsilon>0$. Full support is a key property, because without prior support about the true $p^0_{ \mathcal{L}(\boldsymbol{\mathcal{A}})}$, the posterior cannot concentrate around $p^0_{ \mathcal{L}(\boldsymbol{\mathcal{A}})}$. Theorem \ref{theorem3} provides sufficient conditions on $\Pi_{\nu}$ and the prior for the component-specific edge probability vectors $\Pi_{\pi}$ under which the prior $\Pi$ for $p_{ \mathcal{L}(\boldsymbol{\mathcal{A}})}$, induced through representation (\ref{eq1}),  has full support on  $\mathcal{P}_{2^{V(V-1)/2}}$. 

\begin{theorem}
Let $\Pi$ be the prior induced on the probability mass function $p_{ \mathcal{L}(\boldsymbol{\mathcal{A}})}$ through (\ref{eq1}) and $H^0$ be the number of components required to represent $p^0_{ \mathcal{L}(\boldsymbol{\mathcal{A}})}$ as in \eqref{eq1}. Then  for any $p^0_{ \mathcal{L}(\boldsymbol{\mathcal{A}})} \in  \mathcal{P}_{2^{V(V-1)/2}}$,  $\Pi \{ \mathbb{B}_{\epsilon}(p^0_{ \mathcal{L}(\boldsymbol{\mathcal{A}})})\}>0$ for all $\epsilon>0$ under the following conditions:

(i) $H \ge H^0$ so that $H$ is an upper bound on $H^0$; 

(ii) $\Pi_{\pi}$ has full $L_1$ support on the collections of component-specific edge probability vectors: 

\ \ \ \ \  $\Pi_{\pi}\{\boldsymbol{\pi}^{(1)}, \ldots, \boldsymbol{\pi}^{(H)}: \sum_{h=1}^H \sum_{l=1}^{V(V-1)/2} | \pi^{(h)}_l-\pi^{0(h)}_l |<\epsilon_{\pi}\}>0$, for any collection of edge 

\ \ \ \ \ probability vectors $\{ \boldsymbol{\pi}^{0(1)}, \ldots, \boldsymbol{\pi}^{0(H)}:  \boldsymbol{\pi}^{0(h)}\in (0,1)^{V(V-1)/2}, h=1, \ldots, H \}$ and $\epsilon_{\pi}>0$;

(iii) $\Pi_{\nu} \{ \mathbb{B}_{\epsilon_{\nu}}(\boldsymbol{\nu}^0)\}>0$, for any $\boldsymbol{\nu}^0$ in the probability simplex $\mathcal{P}_H$ and $\epsilon_{\nu}>0$.

\label{theorem3}
\end{theorem}

Lemma  \ref{theorem4} provides sufficient conditions on $\Pi_{Z}$, $\Pi_{X}$, and $\Pi_{\lambda}$ to ensure that the induced prior $\Pi_{\pi}$ through \eqref{eq1.pi} meets condition $(ii)$ in Theorem \ref{theorem3}.

\begin{lemma}
Let $\Pi_{\pi}$ be the prior for the component-specific edge probability vectors induced by $\Pi_{Z}$, $\Pi_{X}$ and $\Pi_{\lambda}$ through  (\ref{eq1.pi}), and denote with $R^0$ a value of $R$ for which Lemma \ref{theorem1} holds, when $p^0_{ \mathcal{L}(\boldsymbol{\mathcal{A}})}$ is factorized as in (\ref{eq1}) with $H^0$ components.  These sufficient conditions imply (ii) in Theorem \ref{theorem3}:

(i) $R \ge R^0$ so that $R$ is an upper bound on $R^0$; 

(ii) $\Pi_{Z}$ has full $L_1$ support on $\Re^{V(V-1)/2}$;

(iii) $\Pi_{X}$ has full $L_1$ support on the space of $V\times R$ real matrices $\Re^{V\times R}$;

(iv) $\Pi_{\lambda}$ has full $L_1$ support on the space of vectors with $R$ real non negative elements $\Re_{\geq 0}^{R}$.

\label{theorem4}
\end{lemma}

These results provide general sufficient conditions on the priors for the components in our factorization under which the induced prior for $p_{ \mathcal{L}(\boldsymbol{\mathcal{A}})}$ has full $L_1$ support.  Full prior support is a key condition of a Bayesian nonparametric model, which also relates to asymptotic behavior of the posterior distribution of $p_{ \mathcal{L}(\boldsymbol{\mathcal{A}})}$. The usual asymptotic focus in the network literature is on the case in which the number of nodes $V \to \infty$ in a single network having a particular structure; see  \citeasnoun**{tang_2013} and \citeasnoun**{suss_2014}. The asymptotic results associated to our methodology provide instead theoretical support  on consistent estimation of the entire population distribution for a network-valued random variable, when the number of networks $n \to \infty$.  Consistently with the type of data we aim to model, we focus on the fixed $V$ case, though it is interesting to develop theory when $V$ increases with $n$.  This is related to a small but growing literature on Bayesian asymptotics in high-dimensional models, but most of the focus has been on substantially simpler models; see \citeasnoun{ghosal_2000} and \citeasnoun{ghosal_2003}.

As the pmf for $ \mathcal{L}(\boldsymbol{\mathcal{A}})$ is characterized by finitely many parameters $p_{ \mathcal{L}(\boldsymbol{\mathcal{A}})}(\boldsymbol{a}), \boldsymbol{a} \in \mathbb{A}_V$, which are all identifiable, full $L_1$ support is sufficient to guarantee that the posterior distribution assigns probability one to any $L_1$ neighborhood of the true data-generating probability mass function as the number of networks $n \to \infty$.  In particular, we have the strong posterior consistency property, 
\begin{eqnarray*}
\lim_{n \to \infty} \Pi\{\mathbb{B}_{\epsilon}(p^0_{ \mathcal{L}(\boldsymbol{\mathcal{A}})}) \mid  \mathcal{L}(\boldsymbol{A}_1), \ldots, \mathcal{L}(\boldsymbol{A}_n)\} = 1, \quad \mbox{for every} \ \epsilon>0,
\end{eqnarray*}
when  $p^0_{ \mathcal{L}(\boldsymbol{\mathcal{A}})}$  is the true probability mass function.

For Theorem \ref{theorem3} and Lemma \ref{theorem4} to hold, we need to choose $H$ and $R$ as upper bounds on $H^0$ and $R^0$.  Then, the priors for the different components in our factorization are chosen to favor automatic adaptation of the model dimension. This is achieved by a double shrinkage prior.    

The first layer of shrinkage collapses out redundant mixture components that are not required to characterize the data. Taking the lead from \citeasnoun{rou_2011}, we let
\begin{eqnarray}
(\nu_1, \ldots, \nu_H) \sim \mbox{Dirichlet}\left(1/H, \ldots, 1/H \right).
\label{eq6}
\end{eqnarray}
In a simpler case involving Gaussian mixtures, \citeasnoun{rou_2011} showed that prior \eqref{eq6} will induce effective deletion of the extra mixture components.  It is an active area of research to extend these asymptotic results on over-fitted mixtures to more general settings. Our empirical results suggest that such efficient deletion of extra components also occurs in our case.  It is straightforward to verify that condition $(iii)$ in Theorem 
 \ref{theorem3} is met under this prior.

The second layer of shrinkage induces collapsing on low-rank structures within each component.  As there are infinitely many positive semidefinite matrices having $\boldsymbol{{D}}^{(h)}$ as lower triangular elements, we are not interested in consistently recovering a true rank for each component-specific deviation vector, but instead look for a prior $\Pi_{\lambda}$ that  shrinks towards a low-rank structure which facilitates borrowing of information in characterizing $\boldsymbol{\pi}^{(h)}$, for $h=1, \ldots, H$ via  \eqref{eq1.pi}. \citeasnoun{bha_2011} address a related goal by developing a shrinkage prior for covariance estimation in  sparse factor models. We adapt their prior to our setting by letting $\boldsymbol{\lambda}^{(h)} \sim \mbox{MIG}(a_1,a_2)$, independently  for $h=1,\ldots, H$, with $\mbox{MIG}(a_1,a_2)$ denoting the Multiplicative Inverse Gamma distribution 
\begin{eqnarray}
\lambda^{(h)}_{r} =\prod_{m=1}^{r}\frac{1}{\vartheta^{(h)}_{m}}, \quad \vartheta^{(h)}_{1}\sim \mbox{Ga}(a_1,1), \ \vartheta^{(h)}_{m\geq2}\sim \mbox{Ga}(a_2,1), \quad r=1, \ldots, R,
\label{eq7}
\end{eqnarray}  
independently for each $h=1, \ldots, H$. Prior \eqref{eq7} adaptively penalizes overparameterized representations for each $\boldsymbol{\pi}^{(h)}$, $h=1, \ldots, H$, by
favoring elements $\lambda^{(h)}_{r}$ to be increasingly concentrated towards zero as $r$ increases, for appropriate $a_2$. Refer to \citeasnoun{bha_2011}  for further details and properties. Additionally $\Pi_{\lambda}$ has a Markovian structure  allowing the joint distribution of $\boldsymbol{\lambda}^{(h)}$  to be factorized as the product of Inverse Gamma distributions. This property facilitates proving Lemma \ref{theorem5}, ensuring that condition $(iv)$ in Lemma \ref{theorem4} is met under this prior choice.
\begin{lemma}
Let $\Pi_{\lambda}$ correspond to the  $\mbox{MIG}(a_1,a_2)$, then $\Pi_{\lambda}$ has full $L_1$ support on $\Re_{\geq 0}^{R}$.
\label{theorem5}
\end{lemma}

Finally, priors $\Pi_{Z}$ and $\Pi_{X}$ are chosen to meet conditions $(ii)$ and $(iii)$, respectively, in Lemma \ref{theorem4}, while favoring simple posterior computation. Consistently with these aims we assume
\begin{eqnarray}
\boldsymbol{Z} \sim \mbox{N}_{V(V-1)/2}(\boldsymbol{\mu}, \boldsymbol{\Sigma}), \quad \boldsymbol{\mu} \in \Re^{V(V-1)/2}, \ \boldsymbol{\Sigma}=\mbox{diag}(\sigma_1^2, \ldots, \sigma^2_{V(V-1)/2}).
\label{eq8}
\end{eqnarray}
Prior $\Pi_{X}$ is defined by assigning independent standard Gaussians  
\begin{eqnarray}
X^{(h)}_{vr}  \sim \mbox{N}(0,1), \quad v=1, \ldots, V, \ \ r=1, \ldots, R, \ \ h=1, \ldots, H.
\label{eq9}
\end{eqnarray}

Beside meeting full prior support conditions, the above settings allow simple derivations for the prior moments of the component-specific log-odds $S^{(h)}_l=Z_l+\sum_{r=1}^R \lambda_r^{(h)}X^{(h)}_{vr}X^{(h)}_{ur}$ for each $h=1, \ldots, H$ and $l=1, \ldots, V(V-1)/2$, with $[v,u]$ denoting the $l$th pair of nodes. Specifically, based on priors \eqref{eq8}--\eqref{eq9} and conditioning on $\boldsymbol{\lambda}^{(h)}$ to highlight their effect, it is easy to show that
\begin{eqnarray}
 \mbox{E}(S^{(h)}_{l} \mid \boldsymbol{\lambda}^{(h)})=\mu_l, \quad \mathrm{var}  (S^{(h)}_{l} \mid \boldsymbol{\lambda}^{(h)})=\sigma_l^{2}+\sum_{r=1}^{R}(\lambda_r^{(h)})^{2}, \quad \mathrm{cov} (S^{(h)}_{l}, S^{(h)}_{l^*}\mid \boldsymbol{\lambda}^{(h)})=0,
 \label{mom}
 \end{eqnarray}
for each $h=1, \ldots, H$, $l=1, \ldots, V(V-1)/2$ and $l^{*}=1, \ldots, V(V-1)/2$ with $l^* \neq l$. The covariance between log-odds in components $h=1, \ldots, H$ and $h^*=1, \ldots, H$ with $h^* \neq h$ is instead
\begin{eqnarray*}
\mathrm{cov} (S^{(h)}_{l}, S^{(h^*)}_{l}\mid \boldsymbol{\lambda}^{(h)}, \boldsymbol{\lambda}^{(h^*)})=\sigma_l^2, \ l=1, \ldots, V(V-1)/2, \quad
\mathrm{cov} (S^{(h)}_{l}, S^{(h^*)}_{l^{*}}\mid \boldsymbol{\lambda}^{(h)}, \boldsymbol{\lambda}^{(h^*)})=0, \ l^* \neq l.
\end{eqnarray*}

{\em A priori} the log-odds of a given edge has the same mean $\mu_l$ in all components with $\sigma_l^2$ controlling the edge-specific portion of variability shared across components as well as the covariance between the log-odds of the same edge in different components. Parameters $\boldsymbol{\lambda}^{(h)}$ add a component-specific portion of variability in the log-odds of each edge.  When the $\boldsymbol{\lambda}^{(h)}$ are all close to zero, the correlation between the log-odds for the same edge across different components is close to one collapsing the model \eqref{eq1}--\eqref{eq1.pi} to a single low-rank representation recalling latent space models for a single network. The prior covariance between the log-odds of different edges is instead zero.

\section{Posterior Computation}
Given priors defined as in equations (\ref{eq6})--(\ref{eq9}), posterior computation for the statistical model having likelihood (\ref{eq1}) with $\boldsymbol{\pi}^{(h)}$ from (\ref{eq1.pi}) is available in a simple form adapting \citeasnoun**{pol_2013} P\'olya-gamma data augmentation for Bayesian logistic regression; see \citeasnoun{cho_2013} for a theoretical justification. Specifically, the proposed Gibbs sampler exploits the graphical representation of our model \eqref{eq1}--\eqref{eq1.pi} outlined in Figure \ref{F_mix} to first allocate each observation $ \mathcal{L}(\boldsymbol{A}_i)$, $i=1, \ldots, n,$ into one of the mixture components and then updates $\boldsymbol{Z}$, $\boldsymbol{X}^{(h)}$, $\boldsymbol{\lambda}^{(h)}$, for $h=1, \ldots, H,$ via Bayesian logistic regression. Detailed steps for implementation are provided in Algorithm \ref{Algorithm_6} below. 
\begin{breakablealgorithm}
\caption{Gibbs sampler for posterior inference in the mixture of low-rank factorization model}
\begin{algorithmic}
\STATE  {{\bf [1] Allocate each network $ \mathcal{L}(\boldsymbol{A}_i)$ to one of the mixture components}}
\vspace{-2pt}
\FOR{$i=1, \ldots, n$}
\vspace{-2pt}
\STATE Sample the class indicator $G_i$ from the discrete distribution with probabilities
\begin{eqnarray*}
\mbox{pr}(G_i=h \mid - )= \frac{ \nu_h \prod_{l=1}^{V(V-1)/2}(\pi_{l}^{(h)})^{ \mathcal{L}(A_i)_l} (1-  \pi^{(h)}_{l})^{1- \mathcal{L}(A_i)_l}}{\sum_{q=1}^{H}\nu_q \prod_{l=1}^{V(V-1)/2}(\pi_{l}^{(q)})^{ \mathcal{L}(A_i)_l} (1-  \pi^{(q)}_{l})^{1- \mathcal{L}(A_i)_l}}, \quad h=1, \ldots, H.
\end{eqnarray*}
\vspace{-15pt}
\ENDFOR
\vspace{-8pt}
\STATE -----------------------------------------------------------------------------------------------------------------------------
\vspace{-8pt}
\STATE  {{\bf [2] Update the mixing probabilities}}
\STATE Sample the mixing probability vector ${\boldsymbol{\nu}}$ from the full conditional Dirichlet $(\nu_1, \ldots, \nu_H) \mid - \sim \mbox{Dirichlet} \{ 1/H+ \sum_{i=1}^{n} \mbox{1}(G_i=1),\ldots , 1/H+ \sum_{i=1}^{n} \mbox{1}(G_i=H)\}$.
\vspace{-8pt}
\STATE -----------------------------------------------------------------------------------------------------------------------------
\vspace{-8pt}
\STATE {\bf {Comment:}} Recalling our generative mechanism in Figure \ref{F_mix}, networks in the same component  $h$ are independent and identically distributed conditionally on the component-specific edge probability vector $\boldsymbol{\pi}^{(h)}$. Hence, to update $\boldsymbol{Z}$, $\boldsymbol{X}^{(h)}$ and $\boldsymbol{\lambda}^{(h)}$, $h=1, \ldots, H$ at each step, it is sufficient to adapt \citeasnoun{pol_2013} P\'olya-gamma data augmentation for aggregated networks $\boldsymbol{Y}^{(1)}, \ldots, \boldsymbol{Y}^{(H)}$, with $\boldsymbol{Y}^{(h)}= \sum_{i:G_i=h} \mathcal{L}(\boldsymbol{A}_i) $, for $h=1, \ldots, H$ and, according to our model 
$$(Y_l^{(h)} \mid  \boldsymbol{Z}, \boldsymbol{X}^{(h)}, \boldsymbol{\lambda}^{(h)}) \sim \mbox{Binom}[n_h,\{1+\exp(-Z_l-\mathcal{L}(X^{(h)}\Lambda^{(h)}X^{(h)\T})_l)\}^{-1}],$$
independently for $l=1, \ldots, V(V-1)/2$ and $h=1, \ldots, H$, with $n_h$ the number of networks in component $h$.  Hence, after the grouping steps, the algorithm proceeds as follows:
\vspace{-8pt}
\STATE -----------------------------------------------------------------------------------------------------------------------------
\vspace{-8pt}
\STATE  {{\bf [3] Data augmentation step via P\'olya-gamma}}
\vspace{-2pt}
\FOR{$h=1, \ldots, H$ and $l=1, \ldots, V(V-1)/2$}
\vspace{-2pt}
\STATE Update the  P\'olya-gamma augmented data from their full conditional P\'olya-gamma
\begin{eqnarray*}
\omega^{(h)}_{l} \mid - \sim \mbox{\small{PG}}\left\{n_h, Z_l+\mathcal{L}(X^{(h)}\Lambda^{(h)} X^{(h)\T})_l\right\},
\end{eqnarray*}
with  $\mbox{\small{PG}}(b,c)$ denoting the  P\'olya-gamma distribution with parameters $b>0$ and $c \in \Re$. 
\vspace{-4pt}
\ENDFOR
\vspace{-8pt}
\STATE -----------------------------------------------------------------------------------------------------------------------------
\vspace{-8pt}
\STATE  {{\bf [4] Update the shared similarity vector}}
\vspace{-2pt}
\STATE Sample the shared similarity vector $\boldsymbol{Z}$ from its Gaussian full conditional
\begin{eqnarray*}
\boldsymbol{Z} \mid - \sim \mbox{N}_{V(V-1)/2}(\boldsymbol{\mu}_Z, \boldsymbol{\Sigma}_Z),
\end{eqnarray*}
with $\boldsymbol{\Sigma}_Z$ diagonal having entries $\sigma^2_{Z_l}=1/(\sigma^{-2}_l+\sum_{h=1}^{H}\omega^{(h)}_{l})$ and $\mu_{Z_l}=\sigma^2_{Z_l}[\sigma_l^{-2}\mu_l+\sum_{h=1}^{H}\{Y_l^{(h)}-n_h/2-\omega^{(h)}_{l}\mathcal{L}(X^{(h)} \Lambda^{(h)} X^{(h)\T})_l\}]$, for each $l=1, \ldots, V(V-1)/2$.
\vspace{-8pt}
\STATE -----------------------------------------------------------------------------------------------------------------------------
\vspace{-8pt}
\STATE {\bf {Comment:}} To maintain conjugacy in sampling the component-specific deviations, we reparameterize the model to update $\boldsymbol{\bar{X}}^{(h)}=\boldsymbol{X}^{(h)}\boldsymbol{\Lambda}^{{(h)}1/2}$ and $\boldsymbol{\Lambda}^{{(h)}}$, $h=1, \ldots, H$. Hence, $\boldsymbol{D}^{(h)}=\mathcal{L}(\boldsymbol{\bar{X}}^{(h)}\boldsymbol{\bar{X}}^{(h)\T})$, and according to our prior specification $\bar{X}_{vr}^{(h)}| \lambda_r^{(h)} \sim \mbox{N}(0, \lambda^{(h)}_r)$ independently for $v=1, \ldots, V$, $r=1, \ldots, R$ and $h=1, \ldots, H$, with independent $\mbox{MIG}(a_1,a_2)$ priors on $\boldsymbol{\lambda}^{(h)}$. 
\vspace{-8pt}
\STATE -----------------------------------------------------------------------------------------------------------------------------
\vspace{-8pt}
\STATE  {{\bf [5] Update the component-specific weighted latent coordinates}}
\vspace{-2pt}
\FOR{$h=1, \ldots, H$}
\vspace{-2pt}
\STATE Block-sample each row of $\boldsymbol{\bar{X}}^{(h)}$, $\boldsymbol{\bar{X}}^{(h)\T}_v=(\bar{X}^{(h)}_{v1}, \ldots, \bar{X}^{(h)}_{vR})$ conditionally on the other parameters and $\boldsymbol{\bar{X}}^{(h)}_{(-v)}$, with $\boldsymbol{\bar{X}}^{(h)}_{(-v)}$ denoting the $(V-1) \times R$ matrix obtained by removing the $v$th row in $\boldsymbol{\bar{X}}^{(h)}$. To do this, we can recast the problem in terms of Bayesian logistic regression 
\vspace{-3pt}
\begin{eqnarray}
\boldsymbol{Y}_{(v)}^{(h)} \sim \mbox{Binom}(n_h,\boldsymbol{\pi}_{(v)}^{(h)}), \quad \mbox{logit}(\boldsymbol{\pi}_{(v)}^{(h)}) =\boldsymbol{Z}_{(v)}+\boldsymbol{\bar{X}}^{(h)}_{(-v)}\boldsymbol{\bar{X}}^{(h)}_v,
\label{eq11}
\end{eqnarray}
\vspace{-3pt}with $\boldsymbol{Y}_{(v)}^{(h)}$ and $\boldsymbol{Z}_{(v)}$ obtained by stacking elements $Y_{l}^{(h)}$ and $Z_l$, respectively, for all $l$ corresponding to pairs $[u,z]$ such that $u=v$ or $z=v$, with $u>z$, and ordered consistently with (\ref{eq11}). Exploiting this formulation, and letting $\boldsymbol{\Omega}_{(v)}^{(h)}$ be the diagonal matrix with the corresponding P\'olya-gamma augmented data, the full conditional is
\begin{eqnarray*}
\boldsymbol{\bar{X}}^{(h)}_v \mid - \sim \mbox{N}_{R}\left\{\left(\boldsymbol{\bar{X}}^{(h)\T}_{(-v)}\boldsymbol{\Omega}_{(v)}^{(h)} \boldsymbol{\bar{X}}^{(h)}_{(-v)}+{\boldsymbol{\Lambda}^{(h)}}^{-1}\right)^{-1} \boldsymbol{\eta}^{(h)}_{v}, \left(\boldsymbol{\bar{X}}^{(h)\T}_{(-v)}\boldsymbol{\Omega}_{(v)}^{(h)} \boldsymbol{\bar{X}}^{(h)}_{(-v)}+{\boldsymbol{\Lambda}^{(h)}}^{-1}\right)^{-1}\right\},
\end{eqnarray*}
with $ \boldsymbol{\eta}^{(h)}_{v}=\boldsymbol{\bar{X}}^{(h)\T}_{(-v)}(\boldsymbol{Y}_{(v)}^{(h)}-\boldsymbol{1}_{V-1} n_h/2-\boldsymbol{\Omega}_{(v)}^{(h)}\boldsymbol{Z}_{(v)})$.
\vspace{-3pt}
\ENDFOR
\vspace{-8pt}
\STATE -----------------------------------------------------------------------------------------------------------------------------
\vspace{-8pt}
\STATE  {{\bf [6] Update the component-specific weight parameters}}
\vspace{-2pt}
\FOR{$h=1, \ldots, H$}
\vspace{-2pt}
\STATE Sample  $ \boldsymbol{\vartheta}^{(h)}=(\vartheta_1^{(h)}, \ldots, \vartheta_R^{(h)})$ characterizing the $\mbox{MIG}(a_1,a_2)$ distribution for $\boldsymbol{\lambda}^{(h)}$ in (\ref{eq7}) 
\vspace{-2pt}
\begin{eqnarray*}
\vartheta_{1}^{(h)} \mid - &\sim& \mbox{Ga} \left\{a_{1}+\frac{V R}{2},1+\frac{1}{2}\sum_{m=1}^{R}\theta_{m}^{(-1)}\sum_{v=1}^{V}(\bar{X}_{vm}^{(h)})^{2}\right\}, \quad \quad \quad \quad \nonumber \\
\vartheta^{(h)}_{r \geq2} \mid - &\sim& \mbox{Ga}\left\{a_{2}+\frac{V (R-r+1)}{2},1+\frac{1}{2}\sum_{m=r}^{R}\theta_{m}^{(-r)}\sum_{v=1}^{V} (\bar{X}_{vm}^{(h)})^{2}\right\},
\end{eqnarray*}
\vspace{-2pt}
where $\theta_{m}^{(-r)}=\prod_{t=1,t\neq r}^{m} \vartheta^{(h)}_{t}$ for $r=1,\ldots,R$.
\vspace{-3pt}
\ENDFOR
\vspace{-8pt}
\STATE -----------------------------------------------------------------------------------------------------------------------------
\vspace{-8pt}
\STATE  {{\bf [7] Update the component-specific edge probability vectors}}
\vspace{-2pt}
\STATE Compute $\boldsymbol{\pi}^{(h)}$ as $\boldsymbol{\pi}^{(h)}=[1+\exp\{-\boldsymbol{Z}-\mathcal{L}(\boldsymbol{\bar{X}}^{(h)}\boldsymbol{\bar{X}}^{(h)\T})\}]^{-1}$, for each $h=1, \ldots, H$.
\vspace{-3pt}
\STATE \vspace{-10pt}
\end{algorithmic}
\label{Algorithm_6}
\end{breakablealgorithm}

Exploiting the MCMC samples for $\boldsymbol{\nu}$ and $\boldsymbol{\pi}^{(1)}, \ldots, \boldsymbol{\pi}^{(H)}$, we can easily obtain samples from the posterior distributions of the expected network summary measures $\bar{\theta}_k$, $k=1, \ldots, K$. In particular, for each MCMC sample of  $\boldsymbol{\nu}$ and $\boldsymbol{\pi}^{(1)}, \ldots, \boldsymbol{\pi}^{(H)}$, we first simulate a sufficiently  large number of networks --- according to  the constructive representation outlined in Figure \ref{F_mix}. For each one of these networks the summary measures of interest are computed and then averaged to obtain an MCMC sample from the posterior distribution of each $\bar{\theta}_k$,  $k=1, \ldots, K$. This provides a tractable strategy to address the inference focus (a).

\subsection{Goodness-of-fit via posterior predictive checks}
Assessing the performance of a statistical model in recovering the generative mechanism underlying the observed data is fundamental to avoid poor characterizations leading to substantially biased inference and conclusions. The importance of this endeavor has motivated an increasing focus on model assessment and comparison for network data; see   \citeasnoun**{hunter_2008}. These procedures evaluate model adequacy by comparing selected network summary measures computed for the observed data, with their distribution --- typically obtained via simulations --- induced by the statistical model of interest. Consistently with these methods,  we assess model adequacy via posterior predictive checks; see e.g. \citeasnoun**{gel_2014}. In particular, we simulate networks from their 
 posterior predictive distribution $p_{\mathcal{L}(\boldsymbol{\mathcal{A}})}(\boldsymbol{a}) \mid  \mathcal{L}(\boldsymbol{A}_1), \ldots, \mathcal{L}(\boldsymbol{A}_n)$ defined as
\begin{eqnarray*}
  \int \sum_{h=1}^{H} \nu_h \prod_{l=1}^{V(V-1)/2} (\pi_{l}^{(h)})^{a_l} (1-  \pi^{(h)}_{l})^{1-a_l} d \Pi_{\pi,\nu}\{\boldsymbol{\nu},\boldsymbol{\pi}^{(1)}, \ldots, \boldsymbol{\pi}^{(H)} \mid  \mathcal{L}(\boldsymbol{A}_1), \ldots, \mathcal{L}(\boldsymbol{A}_n)\},
\label{eq1.pred}
\end{eqnarray*}
for each $\boldsymbol{a} \in \mathbb{A}_V$, with $\Pi_{\pi,\nu}\{\boldsymbol{\nu},\boldsymbol{\pi}^{(1)}, \ldots, \boldsymbol{\pi}^{(H)} \mid  \mathcal{L}(\boldsymbol{A}_1), \ldots, \mathcal{L}(\boldsymbol{A}_n)\}$ representing the posterior distribution of $\boldsymbol{\nu}$ and the component-specific edge probability vectors induced by the posteriors of $\boldsymbol{Z}$, $\boldsymbol{X}^{(h)}$ and $\boldsymbol{\lambda}^{(h)}$, $h=1, \ldots, H$ via \eqref{eq1.pi}. For the simulated networks, a wide set of network measures of interest are computed, obtaining samples from the posterior predictive distributions associated with these measures. If the model is not sufficiently flexible, we expect the network measures computed for the observed data to fall in the tails of their corresponding posterior predictive distributions.  

Beside providing procedures for goodness-of-fit assessments, the posterior predictive distribution for each network summary measure $\theta_k$ represents also a Bayesian estimate  $\hat{p}_{\theta_k}(\theta)$ of the density $p_{\theta_k}(\theta)$ induced by $p_{\mathcal{L}(\boldsymbol{\mathcal{A}})}$ on the network measure  $\theta_k$, providing insights  on the distribution of topological properties across replicates --- according to the inference focus (b).

Although the above integral is not analytically available, it is straightforward to simulate networks from  the posterior predictive distribution of $p_{\mathcal{L}(\boldsymbol{\mathcal{A}})}$ exploiting MCMC samples for $\boldsymbol{\nu}$ and $\boldsymbol{\pi}^{(1)}, \ldots, \boldsymbol{\pi}^{(H)}$, along with the constructive representation of our statistical model outlined in Figure \ref{F_mix} and described in Section 2.

\section{Simulation Study}
We conduct simulation studies to evaluate the performance of our approach in accurately estimating the population distribution of network data, accounting for broad variability in network properties across mixture components.  Of particular interest are community structures \cite{now_2001}, scale freeness \cite{bar_1999}, small-worldness \cite{watts_1998} and classical random graph behaviors  \cite{erd_1959}. Although the latter is overly-restrictive and rarely met in applications, it provides a null model in many network analyses.

\begin{figure}[t]
\centering
\includegraphics[height=4.5cm, width=17cm]{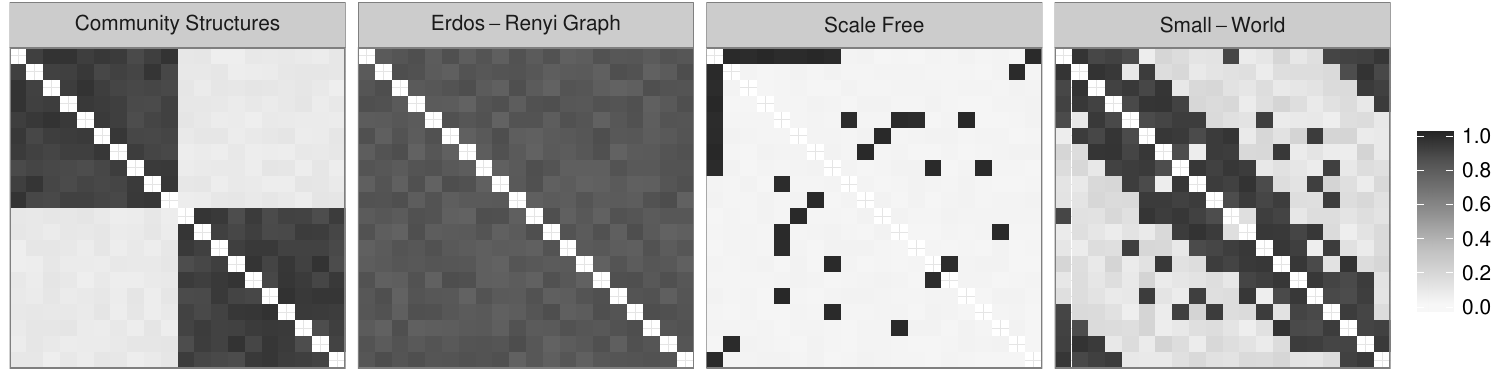}
\caption{\footnotesize{True component-specific edge probability vectors $\boldsymbol{\pi}^{0(h)}$, $h=1, \ldots, 4$ --- rearranged in matrix form.}}
\label{F_simu_true}
\end{figure}

We consider four mixture components and simulate $25$ networks for each component by sampling their edges from conditionally independent Bernoulli random variables given their corresponding component-specific edge probabilities. We focus on networks having $V=20$ nodes to facilitate graphical presentation. Each component-specific edge probability vector is carefully constructed to assign high probability to a subset of network configurations characterized by a specific property. According to  Figure \ref{F_simu_true}, one component is associated with simulated networks characterized by two latent communities. Networks generated under a second component have a behavior similar to \citeasnoun{erd_1959} random graphs. Another component assigns high probability to scale free networks generated under the \citeasnoun{bar_1999} model. Finally networks in the remaining component display small-world properties according to the \citeasnoun{watts_1998}  model.

The goal in defining this challenging simulation scenario is to assess whether our approach can jointly characterize a collection of networks having such broad and widely different properties.  We analyze the simulated data under model (\ref{eq1})--(\ref{eq1.pi}) with priors (\ref{eq6})--(\ref{eq9}). Exploiting results in \eqref{mom}, we consider $\mu_1=\ldots=\mu_{V(V-1)/2}=0$   to obtain priors for each $\boldsymbol{\pi}^{(h)}$ centered on the \citeasnoun{erd_1959} random graph, and let $\sigma_1^2=\ldots=\sigma_{V(V-1)/2}^2=10$ to represent uncertainty in this shared structure. To facilitate automatic adaptation of the latent space dimensions in each component, we consider $a_1=2.5$ and $a_2=3.5$ in the $\mbox{MIG}(a_1,a_2)$ prior.  This enforces adaptive shrinkage for growing $r$, allows component-specific variability in the prior for each  $\boldsymbol{\pi}^{(h)}$ according to  \eqref{mom}, and ensures the existence of the first two moments for the induced priors on elements $D_l^{(h)}$. 

Our approach can be easily modified to learn the hyperparameters from the data via hyperpriors on quantities $\mu_l$, $\sigma_l^2$, for $l=1, \ldots, V(V-1)/2$, $a_1$ and $a_2$. However, we obtained similar results when instead considering other hyperparameter settings, such as $\mu_l \in \{-1,-0.5, 0.5, 1\}$, $\sigma_l^2 \in \{1, 100, 200 \}$ for each $l=1, \ldots, V(V-1)/2$ and $a_1 \in \{5, 10 \}$, $a_2 \in \{5, 10 \}$. Higher values for $a_1$ and $a_2$ are not recommended in inducing priors on $\boldsymbol{\lambda}^{(h)}$ strongly concentrated close to zero, forcing $\boldsymbol{D}^{(h)} \approx 0$. As a result, the component-specific edge probability vectors are forced to be equal, collapsing   \eqref{eq1}--\eqref{eq1.pi} to one low-rank representation recalling latent space models for a single network. This is a similar issue to those encountered in simpler location mixtures of Gaussians when the priors for the component-specific mean parameters are centered on a common hyper-mean, and the hyperprior for the variance hyperparameter concentrates close to zero.  Standard practice avoids these issues by choosing priors that induce moderate variability across 
mixture components.

We generate $5{,}000$ Gibbs iterations, with upper bounds $H=30$ and $R=10$, and set a burn-in of $1{,}000$. Trace-plots suggest this burn-in is sufficient for convergence. We additionally monitor mixing via effective sample sizes for the quantities of interest, with most of these values $\approx 1{,}200$ out of $4{,}000$, providing a good mixing result. The algorithm required 43 minutes to perform posterior computation based on a naive \texttt{R} implementation in a machine with one Intel Core i5 2.3GHz processor and 4GB of RAM. Hence, there are significant margins to improve computational time.

\begin{figure}[t]
\centering
\includegraphics[height=8cm, width=15cm]{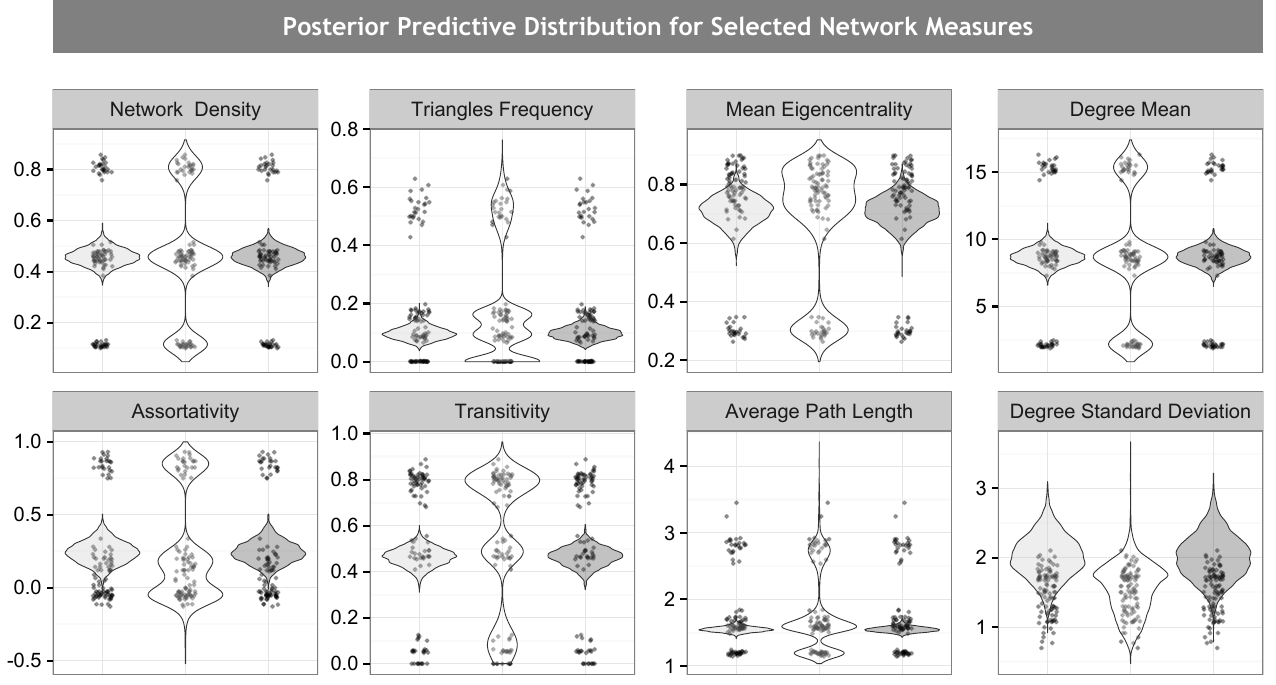}
\caption{\footnotesize{Goodness-of-fit assessments and inference focus (b): For selected network summary measures of interest, violin plots summarizing their posterior predictive distribution $\hat{p}_{\theta_k}(\theta)$, $k=1, \ldots, K$ arising from our procedure (white), the hierarchical latent distance model (light gray) and the hierarchical bilinear model (dark gray). Jittered dots represent the networks summary measures computed for the simulated data.}}
\label{F5}
\end{figure} 

We compare our approach to hierarchical Bayesian formulations \cite**{hand_2007} of the latent distance \cite{hof_2002}  and bilinear \cite{hoff_2005} models.  These models incorporate a flexible factorization for the edge probabilities, with a mixture of Gaussians prior for the latent coordinates improving flexibility.   However, the fundamental issue with these models, and other competitors in the literature, is the assumption of a single edge probability vector $\boldsymbol{\pi}=(\pi_1, \ldots, \pi_{V(V-1)/2})^\T$.   In obtaining posterior samples for $\boldsymbol{\pi}$, we use the  \texttt{R} package  \texttt{latentnet} \cite{kri_2008}, entering the binomial vector $\sum_{i=1}^{n} \mathcal{L}(\boldsymbol{A}_i)$ as the input network since $\sum_{i=1}^{n} \mathcal{L}({A}_i)_l \sim \mbox{Binom}(n, \pi_l)$, for each  $l=1, \ldots, V(V-1)/2$, is a sufficient statistic in these models. 
We use the same MCMC settings as for our model, and the default prior choices for the \texttt{latentnet} routine, obtaining good convergence and mixing performance.  To facilitate comparison with our procedure, we set $R=10$ also in these models and consider four mixture components in the priors for the latent coordinates. Increasing the number of components reduced mixing, without  improving performance. Based on posterior samples for 
$\boldsymbol{\pi}$, inference  and posterior predictive checks proceed as in Sections 2 and 4.

\begin{figure}[t]
\centering
\includegraphics[height=11.5cm, width=12.5cm]{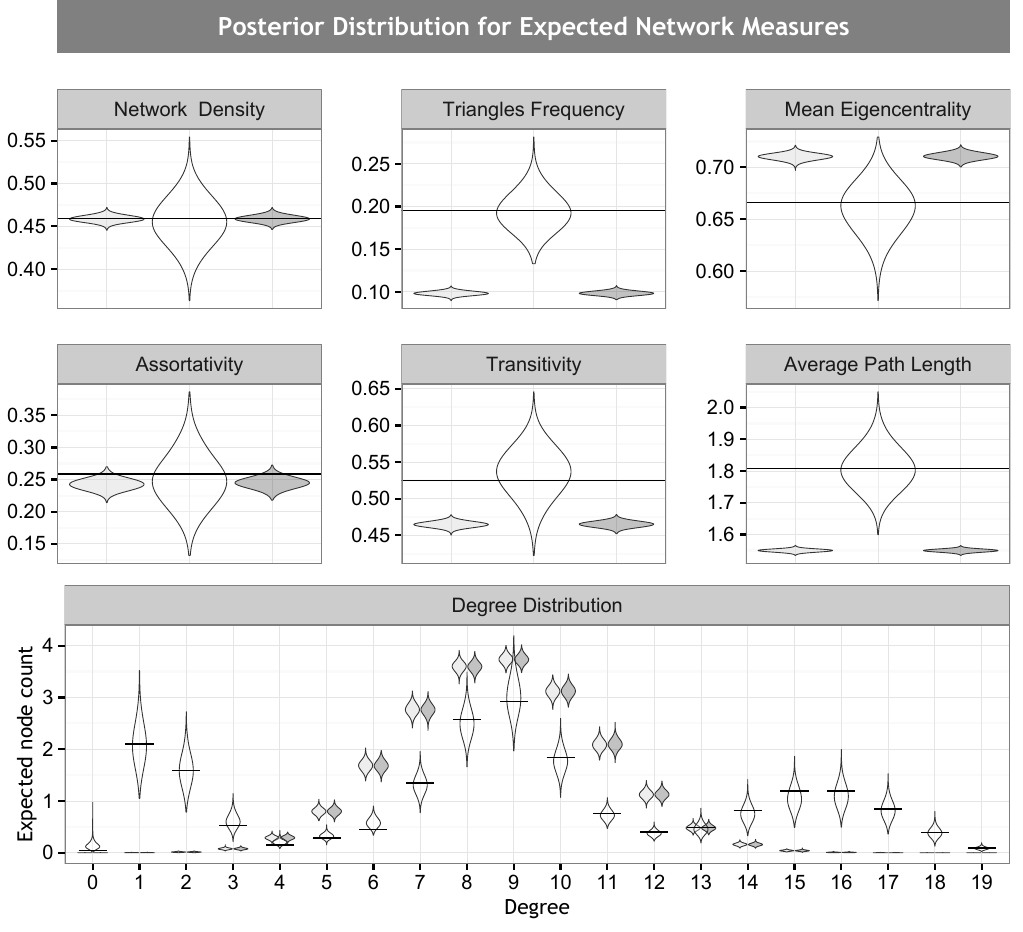}
\caption{\footnotesize{Inference focus (a): Violin plots summarizing the posterior distribution for the expectation of selected network summary measures,  arising from our procedure (white), the hierarchical latent distance model (light gray) and the hierarchical bilinear model (dark gray). Horizontal lines represent the true expected network summary measures. In obtaining posterior samples for each $\bar{\theta}_k$, $k=1, \ldots, K$ we follow the procedure outlined in Section 4, simulating 500 networks for each MCMC sample --- according to the generative mechanism of each model.}}
\label{F4}
\end{figure} 

According to the posterior predictive checks in Figure \ref{F5},  our approach substantially improves flexibility, enhancing performance in uncertainty quantification and providing an adequate representation of the mechanism underlying the simulated networks with respect to a wide variety of different network measures of interest, covering network density, homophily, measures of transitivity,  measures of centrality and statistics summarizing the degree distribution. Accurate predictive performance additionally highlights the flexibility of our low-rank characterization for the mixture components in accommodating a broad range of network structures, while confirming the ability of the shrinkage priors to facilitate automatic adaptation of the model dimension, with  $H=30$ and $R=10$ providing sufficient upper bounds.

\begin{figure}[t]
\centering
\includegraphics[height=5.4cm, width=16.5cm]{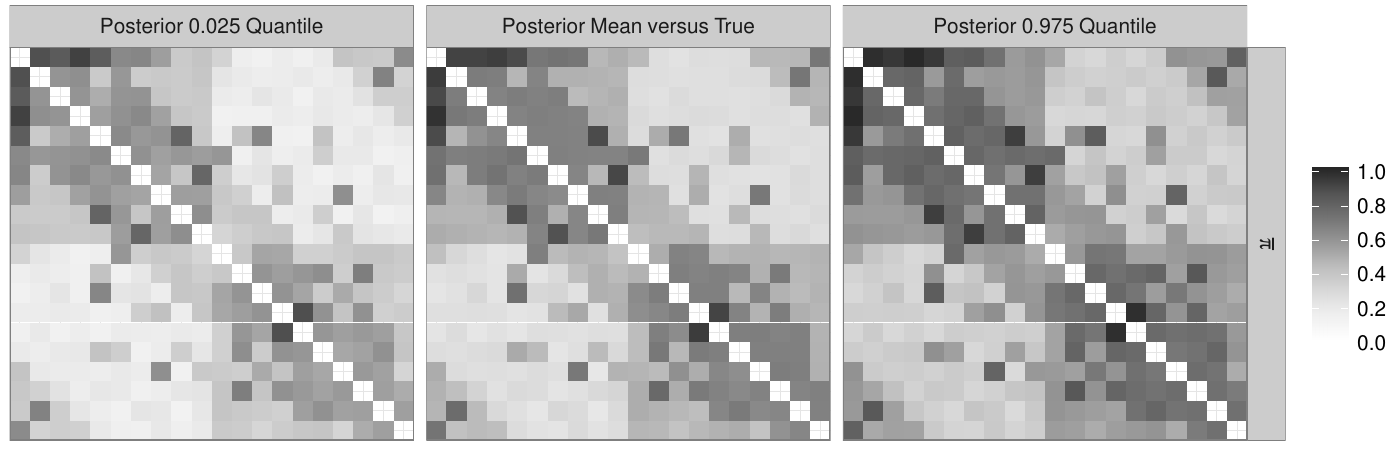}
\caption{\footnotesize{Summary of the posterior distribution for the elements $\bar{\pi}_l$, $l=1, \ldots, V(V-1)/2$ comprising the expected network structure  $\boldsymbol{\bar{\pi}}=\mbox{E}\{\mathcal{L}(\boldsymbol{\mathcal{A}})\}$ --- rearranged in adjacency matrix form. In the middle panel the lower-triangular elements represent the posterior mean $\boldsymbol{\hat{\bar{\pi}}}$ of $\boldsymbol{\bar{\pi}}$, while the upper-triangular elements represent the true expectation $\boldsymbol{\bar{\pi}}^0$}.}
\label{F3}
\end{figure}

The competing approaches produced substantially inferior results ruling out multi-modal patterns in the distribution of the network summary measures and collapsing mass around averaged structures. According to the posterior distribution for the expectation of selected network summary measures in Figure \ref{F4}, this reduced performance in characterizing $p_{ \mathcal{L}(\boldsymbol{\mathcal{A}})}$ substantially affects quality of inference, with the posterior distributions for a wide set of functionals of interest concentrating far from the truth. Due to the superior flexibility of our approach, which is specifically designed for replicated network data, we obtain posterior distributions for the expectation of a wide set of network summary measures of interest accurately centered around the  truth, as shown Figure \ref{F4}. Similar accurate results are obtained when performing posterior inference on the expectation of $\mathcal{L}(\boldsymbol{\mathcal{A}})$ according to Figure \ref{F3}.

To evaluate the performance for increasing $n$ and $V$, Table \ref{tab_2} measures  the concentration of the posterior distributions for the expected network  measures discussed in Figures \ref{F4} and \ref{F3}. Letting $\bar{\theta}_k=\mbox{E}(\theta_k)$, the expectation of the network measure $\theta_k$, we calculate the posterior mean of the absolute difference between $\bar{\theta}_k$ and its true value $\bar{\theta}^0_k$, for a wide set of network measures. 
We consider two scenarios, performing posterior inference with the same settings of our initial simulation study.

The first scenario focuses on the generative mechanism considered in our initial simulation, and evaluates posterior concentration for two different sample sizes $n=40$ and $n=100$, while keeping network size fixed at $V=20$. According to Table \ref{tab_2}, and consistent with our theoretical results on strong posterior consistency, increasing the sample size has the effect of facilitating posterior concentration for all the expected network summary measures.

The second simulation scenario aims instead at providing an empirical evaluation of posterior concentration at increasing network size, for a generative mechanism mimicking --- in a simple version --- the structure and size of our application. Focusing on $V=68$, we simulate $n=40$ networks from our generative mechanism in \eqref{eq1}--\eqref{eq1.pi}, considering four mixture components and two dimensional latent spaces, while  setting the common similarity vector $\boldsymbol{Z}$ equal to the log-odds of the empirical edge probabilities in the brain network application.  As we found low variability across the brains for a wide subset of connections, we force the latent coordinates for half of the nodes to be  zero in all components, while simulating the remaining coordinates from standard Gaussians  with the positive weights $\lambda_r^{(h)}=1$ for each $r$ and $h$. In simulating data for a lower network size $V=34$, we simply hold out from the previous mechanism half of the nodes among those having coordinates equal to zero, and half of the nodes whose coordinates change across the mixture components.

\begin{table}[t]
{\footnotesize{
\begin{tabular}{lccccccc}
& \multicolumn{2}{c}{Scenario 1} &&& \multicolumn{2}{c}{Scenario 2}\\
\hline
&$n=40$, $V=20$ & $n=100$, $V=20$&&& $n=40$, $V=34$ & $n=40$, $V=68$ &\\\hline \hline
Network Density&0.0305& 0.0215&&&0.0042 & 0.0015& \\
Assortativity&0.0495& 0.0322&&& 0.0149 & 0.0082& \\
\hline
Triangles Frequency&0.0263& 0.0173&&& 0.0018 & 0.0006&  \\
Transitivity&0.0446& 0.0268&&& 0.0023 & 0.0012&  \\
\hline
Mean Eigencentrality&0.0248& 0.0189&&& 0.0046 & 0.0030&  \\
Average Path Length&0.0738& 0.0507&&& 0.0051 & 0.0016&  \\
\hline
Degree Distribution&0.2797& 0.1404&&& 0.0461 & 0.0390&  \\
Expectation $\mbox{E}\{\mathcal{L}(\boldsymbol{\mathcal{A}})\}$ &0.0682& 0.0430&&&0.0637 & 0.0635& \\
\hline \hline
\end{tabular}
}}
\renewcommand{\baselinestretch}{1.1} 
\caption{{\footnotesize{Posterior concentration in two different scenarios for varying $n$ and $V$ settings.  For the expected network summary measures considered in Figures   \ref{F4} and \ref{F3}, the table shows the posterior mean of $|\bar{\theta}_k - \bar{\theta}^0_k|$. As the network expectation and the degree distribution are vectors of summary measures, the above expected absolute difference is computed for each element comprising these vectors and then averaged. 
}}}
\renewcommand{\baselinestretch}{1.35} 
 \label{tab_2}
\end{table}

Although the above simulations provide a challenging scenario, with low separation between the mixture components, we obtain good posterior concentration performance for both network sizes --- according to Table \ref{tab_2} --- with this concentration increasing with  $V$. A possible reason for this result is that the underlying latent space dimension remains fixed, and therefore as $V$ increases, borrowing of information across edges via the low-rank representation improves efficiency in modeling the latent coordinates characterizing the mixture components. 

\section{Application to Human Brain Networks}
We  analyze brain connectivity structures in the data set KKI-42 \cite**{land_2011}  available at \texttt{http://openconnecto.me/data/public/MR/archive/}; see \citeasnoun**{cra_2013} for an overview of recent developments in brain imaging technologies and \citeasnoun**{ronc_2013} for details on the construction of brain networks from structural MRI and  Diffusion Tensor Imaging (DTI) scans.

Data are collected for $21$ healthy subjects with no history of neurological disease under a scan-rescan imaging session, so that for each subject two brain network observations are available, for a total of $n=42$. Brain regions are constructed according to the \citeasnoun**{des_2006} atlas, for a total of $V=68$ nodes equally divided in left and right hemisphere. Although recent developments in imaging technologies allow finer parcellations and our algorithms can scale easily to about $V=200$ nodes, the Desikan atlas has been widely utilized in neuroscience studies and hence provides an appealing choice to validate the performance of our model with the respect to available findings.

For each pair of regions and each brain scan, the total number of white matter fibers connecting the two regions in that brain scan is estimated. As a moderate number of these counts is zero,  we focus on the binary adjacency matrices indicating presence of at least one white matter fiber. Fiber tracking pipelines are subject to measurements errors, so there will be inevitably some false positive and false negative edges.  Our model incorporates such measurement errors, and there will be two components of variability in the measured brain networks, one attributable to systematic variability across subjects in their true brain connection structure and one due to measurement errors.  We do not attempt to disambiguate these two components, as this exercise would require some ground truth measurements on actual fibers, which are yet unavailable given current technology.  The proportion of pairs having an edge ranged from a minimum of $0.32$ to a maximum of $0.43$ across scans, with no reason to suspect systematic differences across replicates in measurement errors.

\begin{table}[t]
\begin{center}
{\footnotesize{
\begin{tabular}{lcc}
 \multicolumn{3}{l}{Posterior Distribution for Expected Network Measures} \\ 
\hline
&$a_1=a_2=5$ & $a_1=a_2=10$ \\\hline \hline
Network Density&0.0155& 0.0332 \\
Assortativity&0.0240& 0.0177 \\
\hline
Triangles Frequency&0.0485& 0.0275 \\
Transitivity&0.0908& 0.0457 \\
\hline
Mean Eigencentrality&0.0940& 0.1027  \\
Average Path Length&0.0917& 0.0250  \\
\hline
Degree Distribution&0.0472&0.0641    \\
Expectation $\mbox{E}\{\mathcal{L}(\boldsymbol{\mathcal{A}})\}$ &0.0455& 0.0490 \\
\hline \hline
\end{tabular}
\quad
\begin{tabular}{lcc}
 \multicolumn{3}{l}{Posterior Predictive for Network Measures} \\ 
\hline
&$a_1=a_2=5$ & $a_1=a_2=10$ \\\hline \hline
Network Density&0.0457& 0.0687\\
Assortativity&0.0532& 0.0477   \\
\hline
Triangles Frequency&0.0730& 0.0612  \\
Transitivity&0.0840& 0.0537  \\
\hline
Mean Eigencentrality&0.0307& 0.0447   \\
Average Path Length&0.0480& 0.0577   \\
\hline
Degree Mean&0.0457& 0.0687  \\
Degree Std Deviation&0.0265& 0.0332 \\
\hline \hline
\end{tabular}
}}
\renewcommand{\baselinestretch}{1.1} 
\caption{{\footnotesize{Left table: For the expected network summary measures considered in the inference focus (a), divergence between their posterior distributions when performing inference under different choices of $a_1$ and $a_2$. For each expected network summary measure $\bar{\theta}_k$, the first column displays an estimate of $\mbox{sup}|{F}_{a_1=2.5,a_2=3.5}(\bar{\theta}_k \mid \mbox{data})- {F}_{a_1=5,a_2=5}(\bar{\theta}_k \mid \mbox{data})|$, while the second shows an estimate of $\mbox{sup}|{F}_{a_1=2.5,a_2=3.5}(\bar{\theta}_k \mid \mbox{data})- {F}_{a_1=10,a_2=10}(\bar{\theta}_k \mid \mbox{data})|$, where the generic ${F}_{a_1=a_1^*,a_2=a_2^*}(\bar{\theta}_k \mid \mbox{data})$ denotes the posterior cumulative distribution of $\bar{\theta}_k$, when performing inference with $a_1=a_1^*$, $a_2=a_2^*$. As $\mbox{E}\{\mathcal{L}(\boldsymbol{\mathcal{A}})\}$ and the degree distribution are vectors of summary measures, the above divergence is computed for each element comprising these vectors and then averaged. Right table: Same measures of divergence but focusing on the posterior predictive distribution $\hat{p}_{\theta_k}(\theta)$ for each network measure $\theta_k$, $k=1, \ldots, K$  considered in model checking and in inference focus (b).
}}}
\renewcommand{\baselinestretch}{1.35} 
 \label{tab_3}
\end{center}
\end{table}

Posterior computation under our model is performed as in the simulation study, with the exception of centering the prior for $\boldsymbol{Z}$ on the empirical log-odds of the edges, by letting $\mu_l=\mbox{logit}\{\sum_{i=1}^{42}\mathcal{L}({{A}}_i)_l/42\}$, for each $l=1, \ldots, V(V-1)/2$. We additionally initialize the augmented component indicator variables $G_1, \ldots, G_{42}$ via standard hierarchical cluster analysis applied to the distance matrix between binary vectors $\mathcal{L}(\boldsymbol{A}_1), \ldots, \mathcal{L}(\boldsymbol{A}_{42})$, with the total number of clusters set equal to the upper bound $H=30$. Improved initialization and empirical prior calibration are common practice in Bayesian modeling of networks even in simpler settings focusing on a single network observation and are motivated by the complexity of the non-standard simplex the MCMC routine needs to explore. Refer to   \citeasnoun**{hof_2002}, \citeasnoun**{air_2008},  \citeasnoun{kri_2008} and \citeasnoun**{kri_2009} --- among others --- for notable examples. These finer settings are expected to improve mixing and speed of convergence, while potentially avoiding local modes.

In performing posterior computation, we collect $5{,}000$ Gibbs iterations with a burn-in of $1{,}000$, obtaining good mixing with effective sample sizes $\approx 1{,}100$ out of $4{,}000$. We additionally conducted sensitivity analyses with a specific focus on hyper-parameters $a_1$ and $a_2$ in the  $\mbox{MIG}(a_1,a_2)$ prior for the weight vectors $\boldsymbol{\lambda}^{(h)}$, $h=1, \ldots, H$, controlling shrinkage within each component-specific low-rank factorization. We perform posterior inference also for $a_1=a_2=5$ and $a_1=a_2=10$, and compare the results with those arising from the initial setting of $a_1=2.5$ and $a_2=3.5$. According to the left Table \ref{tab_3}, moderate changes in $a_1$ and $a_2$  do not substantially affect posterior inference on expected network measures. We obtain similar robust results when comparing the posterior predictive distributions for the network summary measures considered in model checking and in the inference focus (b), according to results in the right Table \ref{tab_3}.

\begin{figure}[t]
\centering
\includegraphics[height=8cm, width=15cm]{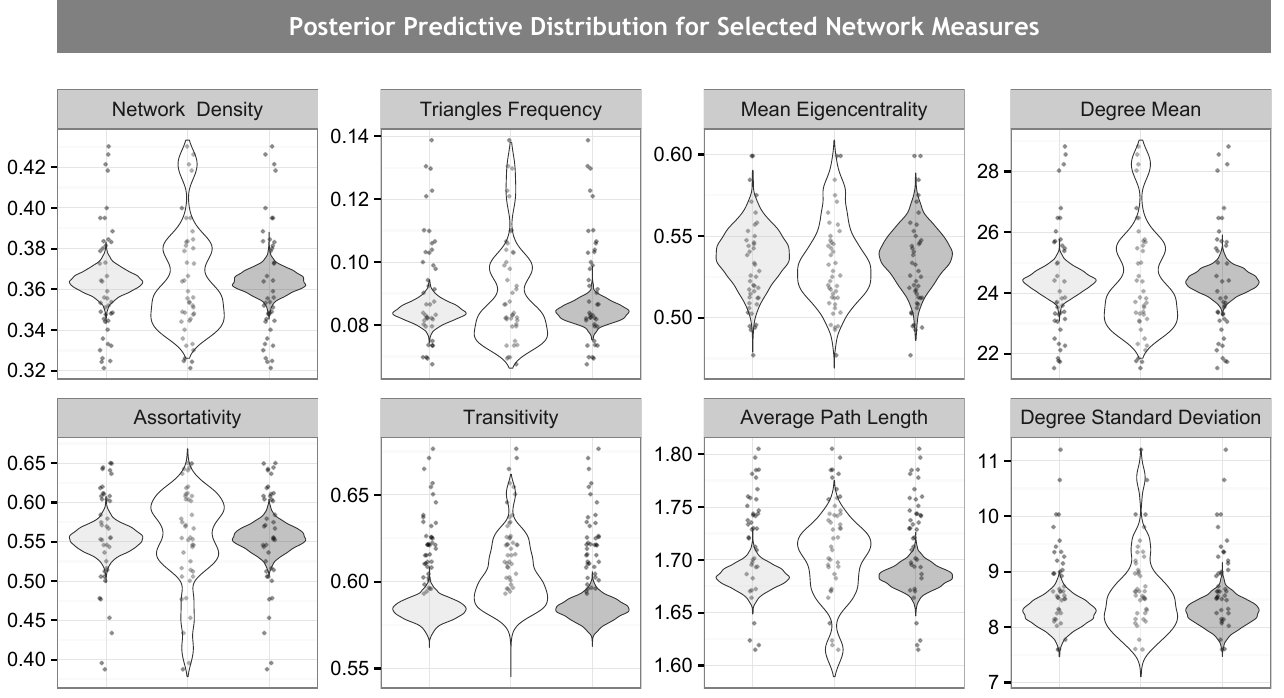}
\caption{\footnotesize{Goodness-of-fit assessments and inference focus (b) in the application: For selected network summary measures of interest, violin plots summarizing their posterior predictive distribution $\hat{p}_{\theta_k}(\theta)$, $k=1, \ldots, K$ arising from our procedure (white), the hierarchical latent distance model (light gray) and the hierarchical bilinear model (dark gray). Jittered dots represent the networks summary measures computed for the observed brains.}}
\label{F9}
\end{figure}

Figure \ref{F9} compares the posterior predictive performance of our approach with the competitors considered in Section 5.  We attempted to improve performance of our competitors by increasing the latent space dimensions and number of mixture components, but did not observed any substantial improvement. 
Consistently with the simulation results, our method substantially improves performance in characterizing the generative mechanism underlying the observed brain networks, correctly estimating multi-modal patterns in the distribution of a wide set of complex network measures and providing accurate uncertainty quantification. The competitors
focus on flexibility in characterizing a single network, and hence provide a restrictive representation of $p_{ \mathcal{L}(\boldsymbol{\mathcal{A}})}$. The impact of this inflexibility is evident from the posterior predictive distributions, which are too concentrated around the averaged structure, while ruling out a wide set of observed network measures.

\begin{figure}[t]
\centering
\includegraphics[height=11.5cm, width=12.5cm]{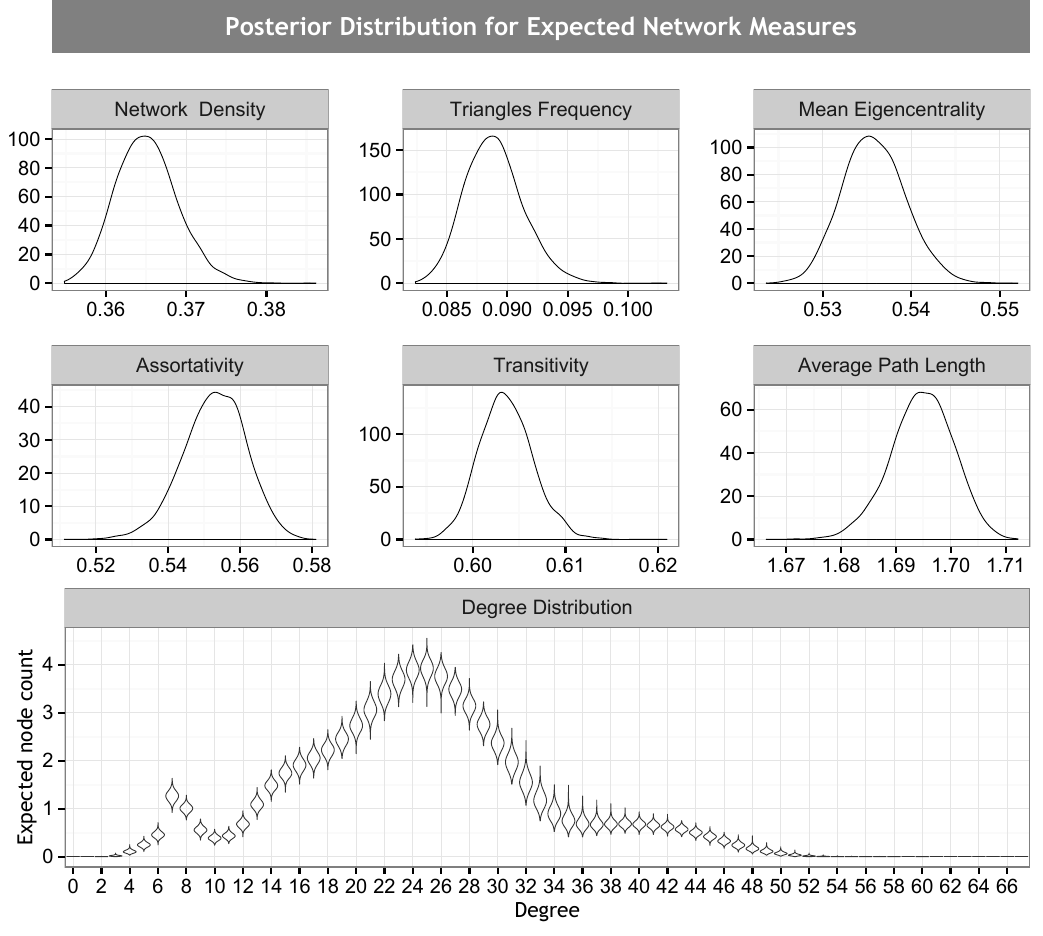}
\caption{\footnotesize{Inference focus (a) in the application: Graphical representation of the posterior distribution for the expectation of selected network summary measures,  arising from our procedure. In obtaining posterior samples for the expected network summary measures $\bar{\theta}_k$, $k=1, \ldots, K$ we follow the procedure outlined in Section 4, simulating 500 networks for each MCMC sample --- according to the generative mechanism of our model.}}
\label{F10}
\end{figure}

Posterior distributions for expected network measures, including network density, homophily by hemisphere, measures of transitivity,  measures of centrality  and degree distribution, are provided for our approach in Figure \ref{F10}.
Consistently with results in Figure \ref{F9} the probabilistic generative mechanism underlying our connectome data is characterized --- on average --- by topological structures indicative of  high hemispheric homophily and  small-world behavior in having small average path length and high transitivity. Small-worldness is confirmed when comparing the above quantities to those obtained from the \citeasnoun{erd_1959} random graph \cite{watts_1998}. Specifically for each network generated to obtain samples from the posterior distribution of the expected network measures in Figure \ref{F10}, we simulate an additional network with the same number of edges considering the \citeasnoun{erd_1959} generative mechanism. We obtain a similar expected average path length  but  lower expected transitivity in the networks simulated under \citeasnoun{erd_1959}, confirming small-worldness according to findings in  \citeasnoun{watts_1998}. 

\begin{figure}[t]
\centering
\includegraphics[height=5.4cm, width=16.5cm]{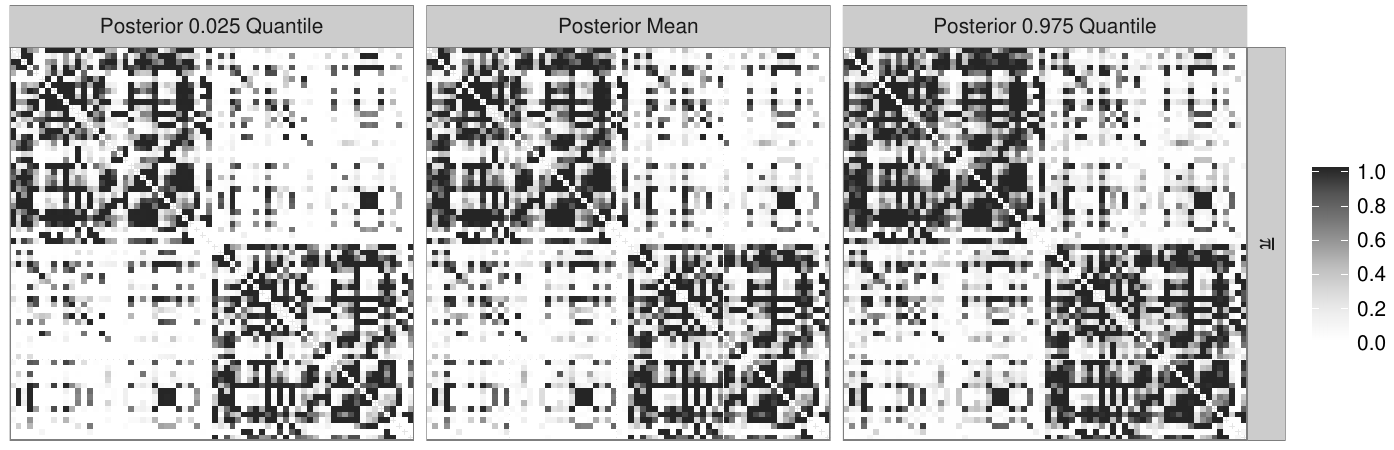}
\caption{\footnotesize{Summary of the posterior distribution for the elements $\bar{\pi}_l$, $l=1, \ldots, V(V-1)/2$ comprising the expected network structure  $\boldsymbol{\bar{\pi}}=\mbox{E}\{\mathcal{L}(\boldsymbol{\mathcal{A}})\}$ --- rearranged in adjacency matrix form.}}
\label{F7}
\end{figure} 

Figure \ref{F7} provides insights into expected network patterns across the different brain scans. The posterior mean for the expected value of $ \mathcal{L}(\boldsymbol{\mathcal{A}})$ highlights a two community structure with intrahemispheric connections more likely than interhemispheric ones, confirming findings on hemispheric homophily in Figures  \ref{F9} and  \ref{F10}. Additionally, several pairs of brain regions have either very high or very low probabilities of connection. For these pairs, there will be little or no differences across monitored brains in the occurrence of edges. The above structure is maintained when considering the $0.025$ and $0.975$ quantiles of the posterior distribution for elements $\bar{\pi}_l$, $l=1, \ldots, V(V-1)/2$, suggesting accurate learning of the expected network structure.

\begin{figure}[t]
\centering
\includegraphics[height=11.5cm, width=12cm]{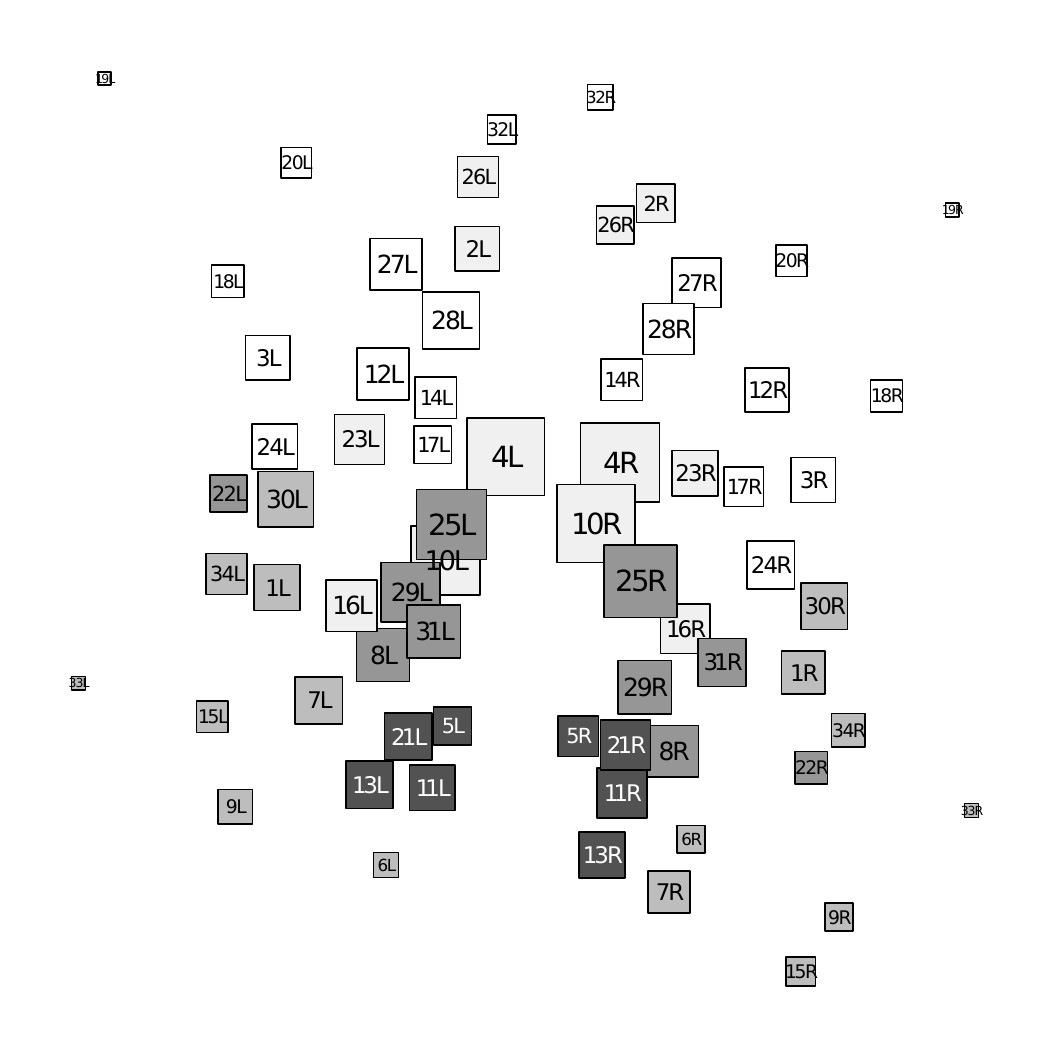}
\vspace{-15pt}
\caption{\footnotesize{Weighted network visualization of the expected brain network architecture with weights given by the posterior mean of $\bar{\boldsymbol{\pi}}$. Node positions are obtained via force-directed placement algorithms. The size of each node  is proportional to its expected degree computed from the estimated ${\bar{\boldsymbol{\pi}}}$. Edges are not displayed to facilitate graphical analysis. Colors define anatomical lobe membership. From lighter to darker gray:  Frontal, limbic, temporal, parietal and occipital lobe. In the node labels, L and R denote regions in the left and right hemisphere, respectively.}}
\label{F8}
\end{figure}

Although our model can be easily extended to accommodate nodal attributes, such as the spatial position and hemisphere membership, we avoid imposing a predefined notion of spatial structure, as one of our goals is to assess whether our model is sufficiently flexible to  learn this structure. Figure  \ref{F8} interestingly shows how this goal is efficiently achieved  by providing a graphical representation of the expected brain network architecture exploiting  the posterior mean of $\bar{\boldsymbol{\pi}}$.  Specifically, node locations are inferred using $\hat{\bar{\boldsymbol{\pi}}}$ via force-directed placement algorithms; see \citeasnoun{fru_1991} for details on network visualization exploiting edge probability information. To highlight hubs in the expected connectivity behavior, we set the size of each node proportional to its expected degree. Nodes having the same color refer to brain regions  belonging to the same lobe according to \citeasnoun**{kang_2012} classification of the Desikan atlas in anatomical lobes. 

In Figure  \ref{F8}, node locations --- inferred from connectivity information --- are interestingly in line with their real spatial coordinates in the brain, while further highlighting the two community structure induced by the hemispheres as well as sub-communities within each hemisphere in line with lobe membership. Figure  \ref{F8} additionally shows how our methodology can provide key insights into hub behaviors. In particular, several regions including corpus callosum --- 4L and 4R --- isthmus --- 10L and 10R --- precunes  --- 25L and 25R --- and superior temporal --- 30L and 30R --- among others --- are characterized by high expected degree and hence represent central anatomical regions for the economy of the brain connectivity architecture. These results are in line  with literature focusing on averaged network structures \cite{bull_2009}. 

We obtained similar insights to those in Figure \ref{F8} when replacing $\hat{\bar{\boldsymbol{\pi}}}$ with the relative edge frequencies $\hat{\boldsymbol{\pi}}=\sum_{i=1}^{42}\mathcal{L}(\boldsymbol{A}_i)/42$ as well as the posterior mean of the edge probability vectors estimated under the hierarchical formulations  of the latent distance and bilinear models. These procedures provide good alternatives and more parsimonious models when the focus is on collapsed network architectures as well as when the replicated networks arise from  probabilistic generative mechanisms concentrating mass around averaged structures. However --- as discussed in Section 5 and recalling the results in Figure \ref{F9} --- these models fall far short of the goal of flexibly modeling more complex and realistic probability mass functions, providing substantially poor results when the focus of inference is on relevant higher-order topological and dependence structures.

\section{Discussion}
We have developed a Bayesian nonparametric approach for modeling the probabilistic generative mechanism associated to a network-valued random variable.  As illustrated through application to human connectome data and simulations, the proposed approach accurately infers expected network structures, while substantially improving state-of-the-art alternatives in adaptively accommodating  the generative mechanism underlying the network-valued data. Our formulation represents a flexible but tractable and efficient procedure for modeling network-valued data, providing procedures for coherent and flexible inference on topological properties and network summary measures, motivating a number of interesting ongoing directions.

One important topic is developing supervised approaches that include predictor variables along with network-valued responses.  For example, behavioral phenotypes, such as creativity, autism, empathy and others, are available in some studies along with brain structural connection network data, with interest focusing on testing for variations in connectivity patterns across such phenotypes \cite{stam_2014}.  Our proposed model can be extended to the supervised case by allowing the mixing probabilities or the latent coordinates to depend on phenotypes.  In such settings, our flexible specification can substantially improve performance, modeling changes in higher-order complex functionals and reducing concerns arising from model miss-specification. 

Additionally, scaling to massive networks is a key issue to deal with the high spatial resolution provided by modern imaging modalities.  Without modification our computational algorithms fail in scaling to very large nodes sets $\mathbb{V}$.  Developing models that exploit sparsity in the network, or avoid sampling through efficient optimization algorithms, provide promising directions. It is additionally worth noticing how our mixture representation \eqref{eq1} naturally incorporates strategies for clustering of networks via posterior inference on the component indicators variables $G_1, \ldots, G_n$. Although this task is beyond the scope of this paper, it is interesting to adapt methodologies for accurate clustering in mixture models \cite**{med_2004,lau_2007}, to our representation.

\section*{Acknowledgements}
This work is graciously supported by the grant CPDA154381/15 of the University of Padova, Italy and by the Defense Advanced Research Projects Agency (DARPA) SIMPLEX program through SPAWAR contract N66001-15-C-4041 and DARPA GRAPHS N66001-14-1-4028.

\section*{Appendix}
\begin{proof}{\bf{of Lemma 2.1}}
To prove the full generality of \eqref{eq1}, note that $p_{ \mathcal{L}(\boldsymbol{\mathcal{A}})}$ is the probability mass function over the cells in a contingency table with the $l$th variable denoting presence or absence of an edge between the $l$th pair of nodes. Hence Lemma 2.1 follows immediately from Theorem 1 of \citeasnoun{dun_2009} with $\boldsymbol{\psi}_h^{(l)} = ( \pi_l^{(h)}, 1-\pi_l^{(h)} )^\T$ for $l=1,\ldots,V(V-1)/2$, as long as any $\boldsymbol{\pi}^{(h)} \in (0,1)^{V(V-1)/2}$ can be represented via \eqref{eq1.pi} for $h=1, \ldots, H$.

Assume without loss of generality $\boldsymbol{Z} = \boldsymbol{0}_{V(V-1)/2}$. Since the logistic mapping is 1-to-1 and continuous it suffices to show that any $\boldsymbol{D}^{(h)} \in \Re^{V(V-1)/2}$ can be expressed as $\boldsymbol{D}^{(h)} = \mathcal{L}( \boldsymbol{X}^{(h)}  \boldsymbol{\Lambda}^{(h)}  \boldsymbol{X}^{(h)\T})$, with $\boldsymbol{X}^{(h)}\in \Re^{V \times R}$ and $\boldsymbol{\Lambda}^{(h)}$ a $R \times R$ diagonal matrix with non-negative entries for each $h=1, \ldots, H$. As there are infinitely many positive semidefinite matrices having lower triangular elements $\boldsymbol{D}^{(h)}$, let $\boldsymbol{\Delta}^{(h)}$ be one of these matrices  such that $\mathcal{L}(\boldsymbol{\Delta}^{(h)}) = \boldsymbol{D}^{(h)}$.  Letting $R^{0(h)}$ the rank of $\boldsymbol{\Delta}^{(h)} = \boldsymbol{\tilde{X}}^{(h)} \boldsymbol{\tilde{\Lambda}}^{(h)}\boldsymbol{ \tilde{X}}^{(h)\T}$, with $\boldsymbol{\tilde{\Lambda}}^{(h)}$ the diagonal matrix with the $R^{0(h)}$ positive eigenvalues of $\boldsymbol{\Delta}^{(h)}$ and $\boldsymbol{\tilde{X}}^{(h)} \in \Re^{V \times R^{0(h)}}$ the matrix with the related eigenvectors, Lemma 2.1 holds after defining $\boldsymbol{X}^{(h)}=\{\boldsymbol{\tilde{X}}^{(h)} \  \boldsymbol{0}_{V \times (R-R^{0(h)})}\}$ and $\boldsymbol{{\Lambda}}^{(h)}$ diagonal, with $\Lambda^{(h)}_{rr}=\tilde{\Lambda}^{(h)}_{rr}$ for $r \leq R^{0(h)}$ and $0$ otherwise.
$\square$
\end{proof}

\begin{proof}{\bf{of Proposition 2.2}}
Focusing on the generic pair $l$ with $l=1, \ldots, V(V-1)/2$, we need show that $\bar{\pi}_l=\sum_{\boldsymbol{a} \in \mathbb{A}_V} a_l p_{ \mathcal{L}(\boldsymbol{\mathcal{A}})}(\boldsymbol{a})=\sum_{h=1}^H \nu_h \pi_l^{(h)}$ for Proposition 2.2 to hold. Under representation \eqref{eq1} for $p_{ \mathcal{L}(\boldsymbol{\mathcal{A}})}$ and letting $\mathbb{A}_{V}^{-l}$ denote the set containing all the possible network configurations for the node pairs except the $l$th one, we can factorize $\bar{\pi}_l$ as 
{\small{
\begin{eqnarray*}
\bar{\pi}_l&=& 1\cdot\{\sum_{\mathbb{A}_{V}^{-l}}  \sum_{h=1}^H \nu_{h} \pi^{(h)}_l \prod_{l^* \neq l} (\pi_{l^*}^{(h)})^{a_{l^*}} (1-  \pi^{(h)}_{l^*})^{1-a_{l^*}}\}{+}0\cdot\{\sum_{\mathbb{A}_{V}^{-l}} \sum_{h=1}^H \nu_{h}(1- \pi^{(h)}_l) \prod_{l^* \neq l} (\pi_{l^*}^{(h)})^{a_{l^*}} (1-  \pi^{(h)}_{l^*})^{1-a_{l^*}}\}\\
&=&\sum_{\mathbb{A}_{V}^{-l}} \sum_{h=1}^H \nu_{h} \pi^{(h)}_l \prod_{l^* \neq l} (\pi_{l^*}^{(h)})^{a_{l^*}} (1-  \pi^{(h)}_{l^*})^{1-a_{l^*}}= \sum_{h=1}^H \nu_{h} \pi^{(h)}_l \sum_{\mathbb{A}_{V}^{-l}} \prod_{l^* \neq l} (\pi_{l^*}^{(h)})^{a_{l^*}} (1-  \pi^{(h)}_{l^*})^{1-a_{l^*}}.
\end{eqnarray*}}}Proposition 2.2 follows after noticing that $\prod_{l^* \neq l} (\pi_{l^*}^{(h)})^{a_{l^*}} (1-  \pi^{(h)}_{l^*})^{1-a_{l^*}}$ is the joint pmf of $V(V-1)/2-1$ independent Bernoulli random variables having joint sample space $\mathbb{A}_{V}^{-l}$ and hence the summation over $\mathbb{A}_{V}^{-l}=\{0,1 \}^{V(V-1)/2-1}$, provides $\sum_{\mathbb{A}_{V}^{-l}} \prod_{l^* \neq l} (\pi_{l^*}^{(h)})^{a_l^{*}} (1-  \pi^{(h)}_{l^*})^{1-a_{l^*}}=1$.
$\square$
\end{proof}

\begin{proof}{\bf{of Theorem 3.1}} As it is always possible to factorize $p^0_{ \mathcal{L}(\boldsymbol{\mathcal{A}})}$ according to \eqref{eq1}, we can express the $L_1$ distance $\sum_{\boldsymbol{a}\in \mathbb{A}_V}| \ p_{ \mathcal{L}(\boldsymbol{\mathcal{A}})}(\boldsymbol{a})-p^0_{ \mathcal{L}(\boldsymbol{\mathcal{A}})}(\boldsymbol{a}) \ |$ between $p_{ \mathcal{L}(\boldsymbol{\mathcal{A}})}$ and $p^0_{ \mathcal{L}(\boldsymbol{\mathcal{A}})}$ as
\begin{eqnarray*}
\sum_{\boldsymbol{a} \in \mathbb{A}_V} \bigg| \ \sum_{h=1}^{H}\nu_h \prod_{l=1}^{V(V-1)/2} (\pi_{l}^{(h)})^{a_{l}} (1-  \pi^{(h)}_{l})^{1-a_{l}}-\sum_{h=1}^{H}\nu^0_h \prod_{l=1}^{V(V-1)/2} (\pi_{l}^{0(h)})^{a_{l}} (1-  \pi^{0(h)}_{l})^{1-a_{l}} \ \bigg|,
\end{eqnarray*}
with vector $\boldsymbol{\nu^0}=(\nu_1^0, \ldots, \nu^0_{H^0}, \boldsymbol{0}_{H-H^0}) \in \mathcal{P}_H$, and $H^0$ the rank of the tensor $p^0_{ \mathcal{L}(\boldsymbol{\mathcal{A}})}$. Hence
\begin{eqnarray*}
\Pi \{\mathbb{B}_{\epsilon}(p^0_{ \mathcal{L}(\boldsymbol{\mathcal{A}})}) \}= \bigintsss \mbox{1}\left(\sum_{\boldsymbol{a} \in \mathbb{A}_V}| \ p_{ \mathcal{L}(\boldsymbol{\mathcal{A}})}(\boldsymbol{a})-p^0_{ \mathcal{L}(\boldsymbol{\mathcal{A}})}(\boldsymbol{a}) \ | <\epsilon\right) d \Pi_{\nu}(\boldsymbol{\nu}) d \Pi_{\pi}(\boldsymbol{\pi}^{(1)}, \ldots, \boldsymbol{\pi}^{(H)}).
\end{eqnarray*}
Following \citeasnoun{dun_2009} and recalling the independence between $\Pi_{\nu}$ and $\Pi_{\pi}$, a sufficient condition for the latter to be strictly positive is that  $\Pi_{\nu}$ has full support on the probability simplex $\mathcal{P}_H$, and $\Pi_{\pi} \{ \mathbb{B}_{\epsilon_{\pi}}(\boldsymbol{\pi}^{0(1)}, \ldots, \boldsymbol{\pi}^{0(H)})\}=\Pi_{\pi}\{\boldsymbol{\pi}^{(1)}, \ldots, \boldsymbol{\pi}^{(H)}: \sum_{h=1}^H \sum_{l=1}^{V(V-1)/2} | \pi^{(h)}_l-\pi^{0(h)}_l |<\epsilon_{\pi}\}>0$, for any collection $\{ \boldsymbol{\pi}^{0(1)}, \ldots, \boldsymbol{\pi}^{0(H)}:  \boldsymbol{\pi}^{0(h)}\in (0,1)^{V(V-1)/2}, h=1, \ldots, H \}$ and $\epsilon_{\pi_h}>0$, which follow from conditions $(iii)$ and $(ii)$ in Theorem 3.1, proving Theorem 3.1. 
$\square$
\end{proof}

\begin{proof}{\bf{of Lemma 3.2}} Letting $\Pi_{S}$ be the prior on the component-specific log-odds vectors induced by $\Pi_Z$, $\Pi_X$ and $\Pi_{\lambda}$ via $\boldsymbol{S}^{(h)}=\boldsymbol{Z}+\mathcal{L}(\boldsymbol{X}^{(h)}\boldsymbol{\Lambda}^{(h)} \boldsymbol{X}^{(h)\T})$, $h=1, \ldots, H$, we first show that for any collection $\{ \boldsymbol{S}^{0(1)}, \ldots, \boldsymbol{S}^{0(H)}:  \boldsymbol{S}^{0(h)}\in \Re^{V(V-1)/2}, h=1, \ldots, H \}$ and $\epsilon_s>0$, $\Pi_S\{\mathbb{B}_{\epsilon_{s}}(\boldsymbol{S}^{0(1)}, \ldots, \boldsymbol{S}^{0(H)}) \}=\Pi_S \{\boldsymbol{S}^{(1)}, \ldots, \boldsymbol{S}^{(H)}: \sum_{h=1}^H\sum_{l=1}^{V(V-1)/2} | \ S^{(h)}_l-S_l^{0(h)} \ |< \epsilon_s\}>0$. Let $R$ be chosen to satisfy condition $(i)$, then according to the proof of Lemma \ref{theorem1}, we can factorize the previous probability as
\begin{eqnarray}
\mbox{pr} \left\{\sum_{h=1}^{H} \sum_{l=1}^{V(V-1)/2} |Z_l-Z^0_l+\mathcal{L}(X^{(h)}\Lambda^{(h)} X^{(h)\T})_l-\mathcal{L}(X^{0(h)}\Lambda^{0(h)}X^{0(h)\T})_l|< \epsilon_s\right\},
\label{teo42.1}
\end{eqnarray}
with $\boldsymbol{\Lambda}^{0(h)}=\mbox{diag}(\boldsymbol{\lambda}^{0(h)})=\mbox{diag}(\lambda_1^{0(h)}, \ldots, \lambda_{R^{0(h)}}^{0(h)}, \boldsymbol{0}_{R-R^{0(h)}})$. Under the independence of $\Pi_Z$, $\Pi_X$ and $\Pi_{\lambda}$, and exploiting the triangle inequality, a lower bound for the previous quantity is
{\small{
 \begin{eqnarray*}
\mbox{pr} \left\{ \sum_{l=1}^{V(V-1)/2} | \ Z_l-Z^0_l \ |<\frac{\epsilon_s}{2H}\right\}\prod_{h=1}^{H} \mbox{pr} \left\{ \sum_{l=1}^{V(V-1)/2}| \ \mathcal{L}(X^{(h)}\Lambda^{(h)} X^{(h)\T})_l-\mathcal{L}(X^{0(h)}\Lambda^{0(h)}X^{0(h)\T})_l \ |< \frac{\epsilon_s}{2H}\right\}.
\end{eqnarray*}}}Hence, (\ref{teo42.1}) is positive if both terms are positive. The positivity of the first term follows from condition $(ii)$ of the Lemma. To prove the positivity of the second term, proof of Lemma \ref{theorem1} ensures that for any $\epsilon_s/(2H)$ there exist infinitely many radii $\epsilon_{X^{(h)}}$, $\epsilon_{\lambda^{(h)}}$, such that $\sum_{v=1}^{V}\sum_{r=1}^{R} | X^{(h)}_{vr} -X^{0(h)}_{vr}|<\epsilon_{X^{(h)}}$ and $\sum_{r=1}^{R} |\lambda^{(h)}_r -\lambda^{0(h)}_r|< \epsilon_{\lambda^{(h)}}$ imply $\sum_{l=1}^{V(V-1)/2}| \ \mathcal{L}(X^{(h)}\Lambda^{(h)} X^{(h)\T})_l-\mathcal{L}(X^{0(h)}\Lambda^{0(h)}X^{0(h)\T})_l \ |< \epsilon_s/(2H)$ for every $h=1, \ldots, H$. Thus to prove the positivity of the second term and recalling the independence between $\Pi_X$ and $\Pi_{\lambda}$, it is sufficient to show that for every $h=1, \ldots, H$ we have $\Pi_{X} \{ \mathbb{B}_{\epsilon_{X^{(h)}}}(\boldsymbol{X}^{0(h)})\}>0$, for any $\boldsymbol{X}^{0(h)} \in \Re^{V\times R}$ and $\epsilon_{X^{(h)}}>0$ and $\Pi_{\lambda} \{ \mathbb{B}_{\epsilon_{\lambda^{(h)}}}(\boldsymbol{\lambda}^{0(h)})\}>0$, for any $\boldsymbol{\lambda}^{0(h)} \in \Re_{\geq 0}^{R}$ and $\epsilon_{\lambda^{(h)}}>0$, representing conditions $(iii)$ and $(iv)$ of the Lemma, respectively.

Let $\pi^{0(h)}_l=\{1+\exp(-S^{0(h)}_l)\}^{-1}$, $l=1, \ldots, V(V-1)/2$, $h=1, \ldots, H$, with $\boldsymbol{S}^{0(h)} \in \Re^{V(V-1)/2}$ factorized as before, and denote with $\Pi_{\pi}$ the prior on the component-specific edge probability vectors, induced by $\Pi_S$ through the 1-to-1 continuous logistic mapping applied element-wise. To conclude the proof of Lemma 3.2 we need to show that $\Pi_{\pi}\{\mathbb{B}_{\epsilon_{\pi}}(\boldsymbol{\pi}^{0(1)}, \ldots, \boldsymbol{\pi}^{0(H)}) \}>0$ given that $\Pi_S\{\mathbb{B}_{\epsilon_{s}}(\boldsymbol{S}^{0(1)}, \ldots, \boldsymbol{S}^{0(H)}) \}>0$ is true. Since the logistic mapping is 1-to-1 element-wise continuous, by the general definition of continuity, for any $\epsilon_{\pi}>0$, there exists an $\epsilon_s>0$, such that
\begin{eqnarray*}
\sum_{h=1}^{H}\sum_{l=1}^{V(V-1)/2} |\ \{1+\exp(-S^{(h)}_l)\}^{-1}- \{1+\exp(-S^{0(h)}_l)\}^{-1} \ |=\sum_{h=1}^H \sum_{l=1}^{V(V-1)/2} | \ \pi^{(h)}_l-\pi^{0(h)}_l \ |<\epsilon_{\pi},
\end{eqnarray*}
for all collections  $\{ \boldsymbol{S}^{(1)}, \ldots, \boldsymbol{S}^{(H)}:  \boldsymbol{S}^{(h)}\in \Re^{V(V-1)/2}, h=1, \ldots, H \}$ such that  $\sum_{h=1}^H\sum_{l=1}^{V(V-1)/2} | S^{(h)}_l-S^{0(h)}_l |< \epsilon_s$. Since we proved that the event $\sum_{h=1}^H\sum_{l=1}^{V(V-1)/2} | S^{(h)}_l-S^{0(h)}_l |< \epsilon_s$ has non-null probability for any $\{ \boldsymbol{S}^{0(1)}, \ldots, \boldsymbol{S}^{0(H)}:  \boldsymbol{S}^{0(h)}\in \Re^{V(V-1)/2}, h=1, \ldots, H \}$, by the continuity of the mapping the same holds for $\sum_{h=1}^H\sum_{l=1}^{V(V-1)/2} | \pi^{(h)}_l-\pi^{0(h)}_l |< \epsilon_\pi$ for any collection $\{ \boldsymbol{\pi}^{0(1)}, \ldots, \boldsymbol{\pi}^{0(H)}:  \boldsymbol{\pi}^{0(h)}\in (0,1)^{V(V-1)/2}, h=1, \ldots, H \}$, concluding the proof. 
$\square$
\end{proof}

\begin{proof}{\bf{of Lemma 3.3}}
Let $\boldsymbol{\lambda}^0$ be a generic vector with $R$ positive elements $\lambda^0_r \in \Re_{> 0}$, $r=1, \ldots, R$. We first show that $\Pi_{\lambda}\{\boldsymbol{\lambda}: \sum_{r=1}^{R} |\lambda_r-\lambda^0_r|< \epsilon_{\lambda}\}>0$, when $\Pi_{\lambda}$ coincides with the $\mbox{MIG}(a_1,a_2)$. Letting $\mathbb{B}_{\epsilon_\lambda}(\boldsymbol{\lambda}^0)=\{\boldsymbol{\lambda}: |\lambda_r-\lambda^0_r|<\epsilon_{\lambda}/R, r=1, \ldots, R \}$ a lower bound for the previous probability is $\Pi_{\lambda}\{\mathbb{B}_{\epsilon_\lambda}(\boldsymbol{\lambda}^0)\}$, and exploiting the Markovian property of the $\mbox{MIG}(a_1,a_2)$ we can factorize this probability as $\int_{\mathbb{B}_{\epsilon_\lambda}(\boldsymbol{\lambda}^0)} f(\lambda_1) \prod_{r=2}^R f(\lambda_r \mid \lambda_{r-1}) d \boldsymbol{\lambda}$, where $f(\lambda_r \mid \lambda_{r-1})$ is the conditional density function of $\lambda_r$ given $\lambda_{r-1}$. 

Hence, the joint $\mbox{MIG}(a_1,a_2)$ prior for $\boldsymbol{\lambda}$ can be factorized as the product of conditional densities with $\lambda_1   \sim \mbox{Inv-Ga}(a_{1},1)$ and $\lambda_{r} \mid  \lambda_{r-1} \sim \mbox{Inv-Ga}(a_{2},\lambda_{r-1})$ for each $r=2, \ldots, R$. Therefore, since the $ \mbox{Inv-Ga}(a,b)$ has full support over $\Re_{>0}$ for any $a>0, b>0$ and provided that by definition $ \lambda_{r-1}>0$ for every $r=2, \ldots, R$, it follows that $\Pi_{\lambda}\{\mathbb{B}_{\epsilon_\lambda}(\boldsymbol{\lambda}^0)\}>0$. This proof holds also for vectors $\boldsymbol{\lambda}^0=(\lambda^0_1, \ldots, \lambda^0_{R^0}, \boldsymbol{0}_{R-R^0})^\T \in \Re^R_{\geq 0}$ with non negative elements as every neighborhood of $\boldsymbol{\lambda}^0$ contains a subset of $\Re^R_{> 0}$ for which prior support has been shown.  This concludes the proof.
$\square$
\end{proof}

\bibliographystyle{ECA_jasa}
\bibliography{example}

\end{document}